\documentclass[aip,reprint,onecolumn,twoside]{revtex4-1}

\usepackage[utf8]{inputenc} 
\usepackage[T1]{fontenc}    
\usepackage{hyperref}       
\usepackage{url}            
\usepackage{booktabs}       
\usepackage{amsfonts}       
\usepackage{nicefrac}       
\usepackage{microtype}      
\usepackage{graphicx}
\usepackage{lipsum}
\usepackage{comment}
\usepackage[caption=false]{subfig}
\usepackage{textcomp}
\usepackage{soul}           
\usepackage{amsmath}
\usepackage{listings}
\usepackage{chngcntr}
\usepackage{enumitem}
\usepackage[table,xcdraw]{xcolor}
\usepackage{booktabs}
\usepackage{multirow}
\usepackage{hhline}
\usepackage{outlines}
\usepackage{empheq}
\usepackage{sidecap}
\usepackage{physics}

\definecolor{codegreen}{rgb}{0,0.6,0}
\definecolor{codegray}{rgb}{0.5,0.5,0.5}
\definecolor{codepurple}{rgb}{0.58,0,0.82}
\definecolor{backcolour}{rgb}{0.95,0.95,0.92}

\lstdefinestyle{mystyle}{
    backgroundcolor=\color{backcolour},   
    commentstyle=\color{codegreen},
    keywordstyle=\color{magenta},
    numberstyle=\tiny\color{codegray},
    stringstyle=\color{codepurple},
    basicstyle=\ttfamily\footnotesize,
    breakatwhitespace=false,         
    breaklines=true,                 
    captionpos=b,                    
    keepspaces=true,                 
    numbers=left,                    
    numbersep=5pt,                  
    showspaces=false,                
    showstringspaces=false,
    showtabs=false,                  
    tabsize=2
}
\let\oldtextbf=\textbf
\renewcommand*{\textbf}[1]{\ifmmode\mathbf{#1}\else\oldtextbf{#1}\fi}
\renewcommand*{\phi}[0]{\varphi}

\newcommand{\sansbf}[1]{\textsf{\textbf{\small {#1}}}}
\newcommand{\paragraphtitle}[1]{\vspace{0.5em}\noindent\textsf{\textbf{\small {#1}}}}
\newcommand{\mybox}[1]{\noindent\fbox{\parbox{\textwidth}{{#1}}}\vspace{0.5em}}
\newcommand{\linkbox}[1]{\framebox{\scriptsize \textsf{#1}}\,}

\makeatletter
\newcommand{\tocfill}{\cleaders\hbox{$\m@th \mkern\@dotsep mu . \mkern\@dotsep mu$}\hfill}
\makeatother

\newenvironment{abbreviations}{\twocolumngrid\footnotesize\begin{list}{}
    {%
    \setlength{\itemsep}{0.1pt}}
    }
{\end{list}\onecolumngrid}
\usepackage{titlesec}
\titleformat{\section}  
  {\fontsize{14}{15}\sffamily\bfseries} 
  {\thesection} 
  {1em} 
  {} 
  [] 

\titleformat{\subsection}  
  {\fontsize{12}{14}\sffamily\bfseries} 
  {\thesubsection} 
  {1em} 
  {} 
  [] 

\titleformat{\subsubsection}  
  {\fontsize{11}{13}\sffamily\bfseries} 
  {\thesubsubsection} 
  {1em} 
  {} 
  [] 

\setlist[itemize]{leftmargin=*}
\setlist[enumerate]{leftmargin=*}

\usepackage{fancyhdr}
\fancypagestyle{plain}{%
    \fancyhead{} 
    \fancyfoot[C]{\thepage}
}
\begin{document}

\setlength{\abovedisplayskip}{4pt}
\setlength{\belowdisplayskip}{4pt}

\title{{\Large Advanced simulations with PLUMED:} \\{\LARGE OPES and Machine Learning Collective Variables}\vspace{0.5em}} 

\author{Enrico Trizio}
\affiliation{Atomistic Simulations, Italian Institute of Technology, 16156 Genova, Italy}

\author{Andrea Rizzi}
\affiliation{Atomistic Simulations, Italian Institute of Technology, 16156 Genova, Italy}
\affiliation{Computational Biomedicine, Institute of Advanced Simulations IAS-5; Institute for Neuroscience and Medicine
INM-9, Forschungszentrum Jülich GmbH, Jülich 52428, Germany}

\author{Pablo M. Piaggi}
\affiliation{CIC nanoGUNE BRTA, 
20018 Donostia-San Sebastián, Spain}
\affiliation{Ikerbasque, Basque Foundation for Science, 48013 Bilbao, Spain}

\author{Michele Invernizzi}
\affiliation{Peptone Ltd., 
London, NW1 1JD, United Kingdom}

\author{Luigi Bonati}
\email[Corresponding author: Luigi Bonati]{ (luigi.bonati@iit.it)}
\affiliation{Atomistic Simulations, Italian Institute of Technology, 16156 Genova, Italy}

\begin{abstract}
Many biological processes occur on time scales longer than those accessible to molecular dynamics simulations. Identifying collective variables (CVs) and introducing an external potential to accelerate them is a popular approach to address this problem. 
In particular, \verb|PLUMED| is a community-developed library that implements several methods for CV-based enhanced sampling.
This chapter discusses two recent developments that have gained popularity in recent years. The first is the On-the-fly Probability Enhanced Sampling (OPES) method as a biasing scheme. This provides a unified approach to enhanced sampling able to cover many different scenarios: from free energy convergence to the discovery of metastable states, from rate calculation to generalized ensemble simulation. 
The second development concerns the use of machine learning (ML) approaches to determine CVs by learning the relevant variables directly from simulation data. The construction of these variables is facilitated by the \verb|mlcolvar| library, which allows them to be optimized in Python and then used to enhance sampling thanks to a native interface inside PLUMED. 
For each of these methods, in addition to a brief introduction, we provide guidelines, practical suggestions and point to examples from the literature to facilitate their use in the study of the process of interest.
\end{abstract}
\maketitle

\addtocontents{toc}{\vspace{-1.em}}

 \begin{figure}[h!]
     \centering
\includegraphics[width=0.55\linewidth]{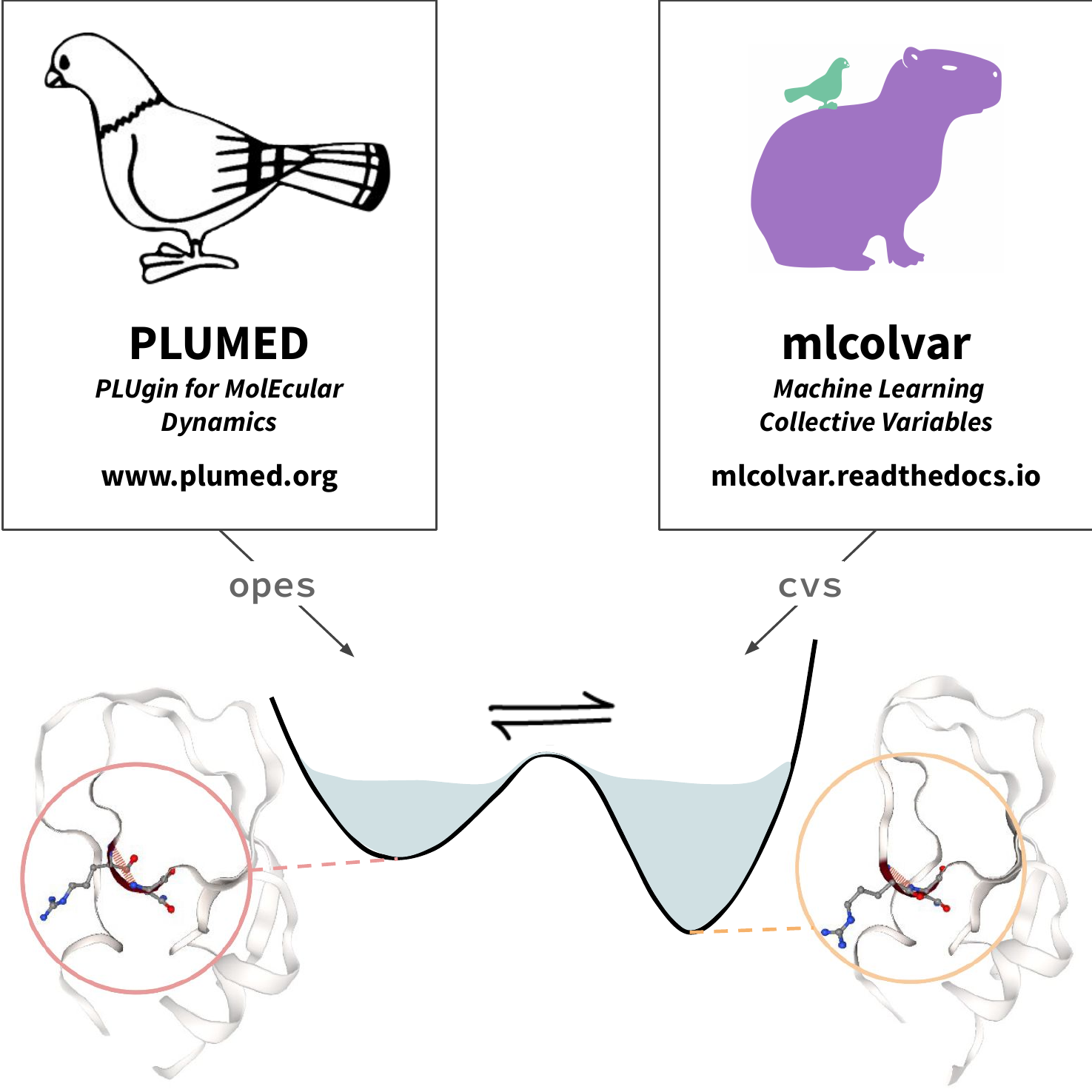} 
\end{figure}

\clearpage
\tableofcontents
\subsection*{List of abbreviations}
\begin{abbreviations}
    \item[AE]           Autoencoder 
    \item[CV]           Collective variable
    \item[ESS]          Effective sample size
    \item[FES]          Free energy surface
    \item[FFNN]         Feedforward neural network
    \item[GNN]          Graph neural network
    \item[KDE]          Kernel density estimation
    \item[LASSO]        Least absolute shrinkage and selection operator
    \item[LDA]          Linear discriminant analysis 
    \item[ML]           Machine learning
    \item[MLCV]         Machine learning collective variable
    \item[MD]           Molecular dynamics
    \item[MetaD]        Metadynamics
    \item[NN]           Neural network
    \item[OPES]         On-the-fly probability enhanced sampling
    \item[PCA]          Principal component analysis
    \item[RMSD]         Root mean squared deviation
    \item[TDA]          Targeted discriminant analysis
    \item[TICA]         Time-lagged independent component analysis
    \item[TPI]          Transition path informed 
    \item[TPS]          Transition path sampling
    \item[TS]           Transition state
    \item[WT-MetaD]     Well-tempered metadynamics
\end{abbreviations}

\clearpage
\section{Introduction}
\paragraphtitle{Rare Events and Enhanced Sampling}. Molecular dynamics (MD) is a numerical technique that allows to simulate the time evolution of atomistic systems~\cite{frenkel2001understanding}. 
It provides access not only to the equilibrium properties of the system under study but also to the kinetics and microscopic transition mechanisms between its different metastable states. 
For example, we can investigate the mechanism behind the folding of a protein or the thermodynamics of binding of a ligand to a receptor macromolecule.
However, the study of these processes is far from trivial, because metastable states are very often separated by large free-energy barriers.
As a result, transitions between them occur on time scales longer than what we can simulate in practice with standard MD simulations. 
To circumvent this so-called rare event problem, many enhanced sampling methods have been proposed, which can be grouped into three categories~\cite{pietrucci2017strategies}. 
The first family of techniques relies on the identification of collective variables (CVs) and introduces an external potential to enhance their fluctuations, thus reducing the free energy barriers along them, as in the case of Umbrella Sampling~\cite{torrie1977nonphysical, mezei1987adaptive}, Metadynamics~\cite{laio2002escaping, barducci2008welltempered} and Variationally Enhanced Sampling~\cite{valsson2014variational,bonati2019neural}.
In the second group, the equilibrium distribution is broadened in a more generic way, as in the case of replica exchange molecular dynamics~\cite{sugita1999replicaexchange} or generalized ensembles~\cite{lyubartsev1992new,invernizzi2020unified}.
Finally, the third category aims at sampling reactive paths by doing, for example, a Monte Carlo simulation in trajectory space\cite{bolhuis2002transition}. 
We refer the reader to the review by Henin \textit{et. al.} for a comprehensive overview of enhanced sampling methods~\cite{Henin2022enhanced}. 
In what follows, we will focus mostly on the first family of CV-based methods while highlighting some connections and synergistic interactions with the other families. 

\paragraphtitle{Recent developments in CV-based enhanced sampling}. 
This chapter assumes some basics about enhanced sampling based on collective variables; for readers less familiar with the theoretical underpinnings of such methods, we suggest starting from the reviews by Valsson \textit{et. al.}~\cite{valsson2016enhancing} and by Bussi and Laio~\cite{bussi2020using}.
Here, we would like to focus on some advanced methods that have been proposed in recent years, which cover the two main ingredients of enhanced sampling methods, i.e., the identification of CVs and the scheme for applying the external potential.
Both are essential for successfully obtaining free energy profiles and reconstructing transition mechanisms. 
From the point of view of biasing schemes, we will focus on OPES~\cite{invernizzi2020rethinking, invernizzi2022exploration, invernizzi2020unified} (on-the-fly probability enhanced sampling), an evolution of the well-known Metadynamics~\cite{laio2002escaping, barducci2008welltempered} technique. 
Then, we will discuss machine learning (ML) approaches for a data-driven design of CVs directly from simulations. 
Both these approaches have found widespread use in recent years and opened new avenues in enhanced sampling calculations. 
As we discuss these advances, we want to provide a practical guide to help readers understand the main ingredients for using them in their research through the available software.
In particular, this is centered on the \verb|PLUMED| plugin for free energy and enhanced sampling calculations, assisted by the \verb|mlcolvar| library for the data-driven construction of CVs, which we briefly discuss in the following paragraphs.

\paragraphtitle{Software for enhanced sampling}. PLUMED~\cite{tribello2014plumed, plumed2019promoting} is an open-source, community-developed library that offers a comprehensive collection of enhanced sampling methods, free-energy techniques, and tools for analyzing MD simulation data.
It is a very versatile tool, as it is compatible with many popular MD engines (AMBER~\cite{amber2013}, GROMACS~\cite{abraham2015gromacs}, LAMMPS~\cite{lammps}, NAMD~\cite{phillips2020namd}, CP2K~\cite{kuhne2020cp2k}, OPENMM~\cite{eastman2017openmm}, to name a few) and provides ingredients to easily construct collective variables to describe complex processes in physics, chemistry, materials science and, certainly, biology. 
To familiarize with this tool, extensive resources are available.
In particular, we suggest starting with the book chapter “Analyzing and Biasing Simulations with PLUMED”~\cite{bussi2019analyzing} from the previous edition of \textit{Biomolecular Simulations: Methods and Protocols} which covers the basics of PLUMED both for analysis and for sampling.
Other important resources are:
    \begin{itemize}
        \itemsep0em 
        \item PLUMED-TUTORIALS~\footnote{\url{https://www.plumed-tutorials.org/}}, a community-driven learning ecosystem, featuring also the PLUMED-MASTERCLASS series with YouTube-recorded lectures and hands-on exercise corrections. These tutorials cover both the basic syntax as well as many advanced features contributed by the community. We will point to some of them with the box:        
        \linkbox{TUTORIAL:\href{https://www.plumed-tutorials.org}{XX.YYY}}.
        \item PLUMED-NEST~\footnote{\url{https://www.plumed-nest.org/}}, a repository of input data to reproduce published results of PLUMED-based simulation. We will refer to these entries  \linkbox{NEST-\href{https://www.plumed-nest.org/}{XX.YYY}} to point to relevant examples of how the methodologies discussed are used in practice to study different types of systems.
    \end{itemize} 

From a software perspective, \verb|PLUMED| consists of a core part plus many modules contributed by the community to expand its functionality.
In particular, the topics we cover in this chapter require two modules: \verb|opes|, regarding enhanced sampling, and \verb|pytorch|, which implements an interface to load into  \verb|PLUMED| CV models optimized in Python with the machine learning library PyTorch~\cite{paszke2019pytorch} and compiled with the TorchScript language. 

\paragraphtitle{Machine Learning Collective Variables}.  \verb|mlcolvar|~\cite{bonati2023mlcolvar} is a Python library based on PyTorch~\cite{paszke2019pytorch} aimed at facilitating the construction of data-driven CVs for enhanced sampling.
This library has several purposes: first, it allows for the use (or easy implementation) of several CVs proposed in the literature. 
It is structured in a very modular way so that it is easy to assemble new CVs using different combinations of, for example, architectures and loss functions. 
This is intended to simplify the development of new approaches and the contamination between them. 
Last but not least, it is designed to work with PLUMED, enabling a straightforward deployment of ML-based CVs within the enhanced sampling method of choice and the preferred MD code. 
All the steps discussed in this chapter, from the training to the interpretation of the CVs, are indeed implemented in \verb|mlcolvar|, and we will refer the reader to practical Jupyter Notebook tutorials from the documentation with the box \linkbox{MLCOLVAR-\href{https://mlcolvar.readthedocs.io/en/latest/tutorials.html}{DOC}}.

\section{Enhancing sampling with OPES}
\label{sec:OPES}

As mentioned in the introduction, OPES was created as an evolution of the popular Metadynamics enhanced sampling method, shifting the focus from constructing the bias potential to choosing the probability distribution to be sampled. OPES can be used to obtain the same type of sampling as Metadynamics in a more robust way, also requiring fewer parameters to set. 
Moreover, the OPES framework has proven to be extremely flexible, providing a unified approach that can achieve multiple goals, ranging from exploration (see Sec.~\ref{sec:opes_explore}) to convergence of free energies (Sec.~\ref{subsec:OPES-metad}), rate calculation (Sec.~\ref{sec:opes_flooding}), and generalized ensemble simulations (Sec.~\ref{sec:opes_variants}).
To fully appreciate the capabilities of this tool, we begin this section with a brief discussion of enhanced sampling as a method for altering the probability distribution while still being able to recover information about the original physical system.

\subsection{Free energies and probability distributions}
\label{sec:free_energies}

\begin{SCfigure}
    \centering
    \includegraphics[width=0.65\linewidth]{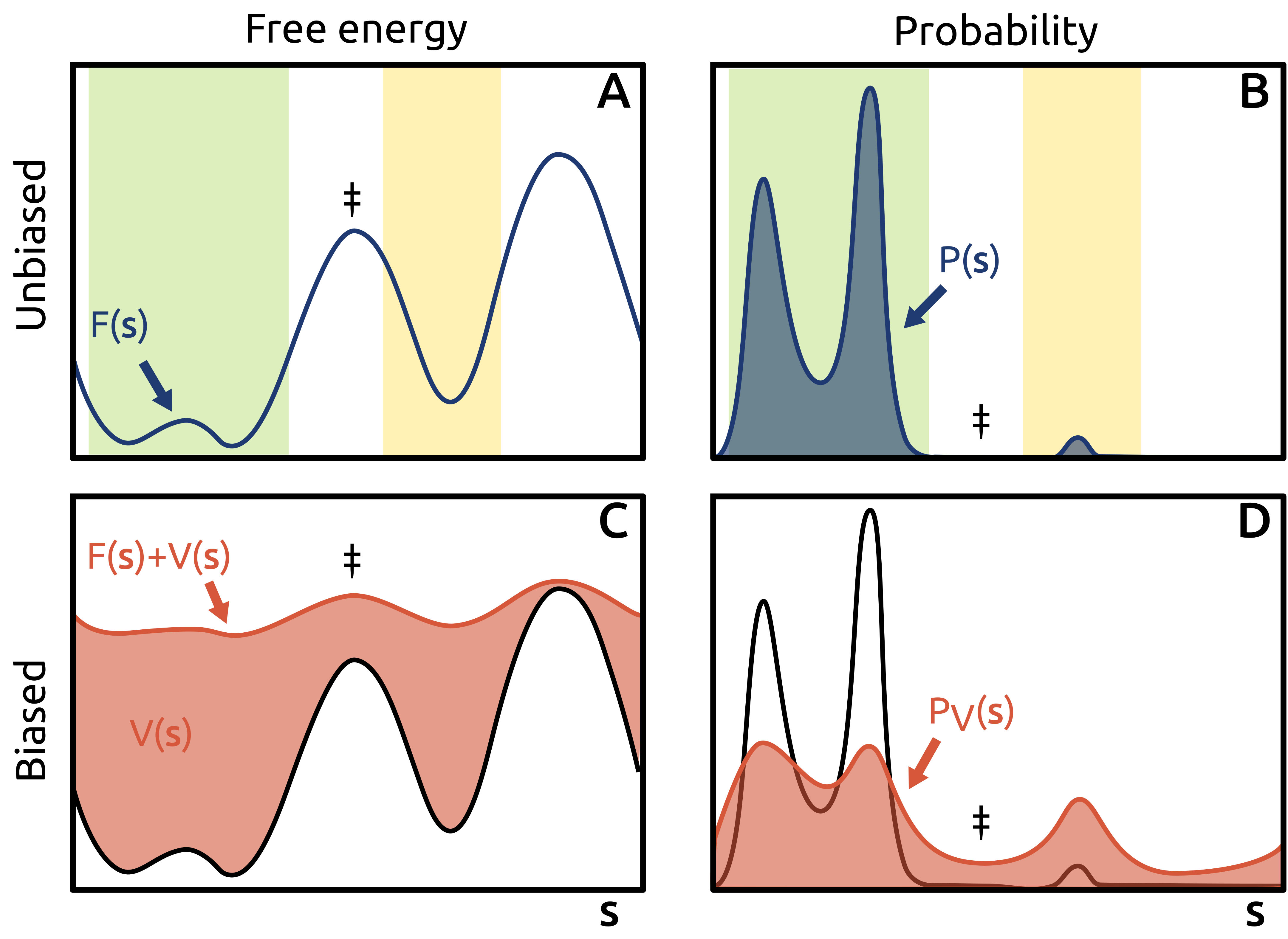}
    \caption{Schematical visualization of the relation between free energy (\textbf{A}, \textbf{C}) and probability distribution (\textbf{B}, \textbf{D}) along a collective variable $\textbf{s}$.
    For an unbiased system with free energy $F(\textbf{s})$ (top row), the probability distribution is strongly peaked on the metastable states, highlighted by the green and yellow shaded box, and almost negligible in correspondence to the free energy barriers associated with the transition state ($\ddagger$).
    If a proper bias potential $V(\textbf{s})$ is applied to the system (bottom row), the relative free energy barriers are lowered, and, as a consequence, the probability distribution is broadened, also covering the transition state region.  }
    \label{fig:probability-potential}
\end{SCfigure}

\paragraphtitle{Probability and free energy profiles.} 
It is well known that many natural processes of interest are rare events when compared to the time scales typically accessible to standard MD, making their simulation challenging.
This is due to the presence of large free energy barriers separating the metastable states of the system, for example, the folded and unfolded conformations of a protein. 
From another point of view, the same concept can be described in terms of the probability of observing each configuration in the different regions of the phase space in our simulations. 
In a canonical ensemble, this probability is related to the potential energy $U(\textbf{x})$ of the system through the well-known Boltzmann distribution
\begin{equation}\label{eq:boltzmann-distribution}
p(\textbf{x}) = \frac{e^{-\beta U(\textbf{x})}}{Z} \quad \text{with} \quad Z = \int \dd\textbf{x}\ e^{-\beta U(\textbf{x})}
\end{equation}
where $\beta=1/k_BT$ is the inverse temperature ($k_B$ the Boltzmann constant), and $Z$ is a normalization constant called configurational partition function. 
The samples from an MD simulation in the canonical ensemble will be indeed distributed according to Eq.~\ref{eq:boltzmann-distribution}.
Since this probability lives in the full $3N$ dimensional space, where $N$ is the number of atoms, it is convenient to understand the system in terms of a smaller set of degrees of freedom, which are referred to as \textit{collective variables} (CVs). 
These CVs are, in general, functions of the atomic coordinates of the system $\textbf{s} = \textbf{s}(\textbf{x})$, and, in the same spirit of reaction coordinates in chemistry or order parameters in physics, are intended to capture the progress of the molecular process under study.
With one or a few CVs at hand, one can define the probability of observing a given value of {$\textbf{s}$} during the simulation as the marginal distribution of $p(\textbf{x})$ (see Fig.~\ref{fig:probability-potential}) along $\textbf{s}$
\begin{equation}\label{eq:density-cv}
    p(\mathbf{s}) = \int \dd\textbf{x}\ \delta(\textbf{s}-\textbf{s}(\textbf{x})) p(\mathbf{x}).
\end{equation}
From Fig.~\ref{fig:probability-potential}B, it can be seen that in the case that in rare events, this probability is multimodal, with peaks on metastable states separated by regions where the probability is almost negligible. 
Consequently, as systems must traverse these unfavorable configurations to move from one state to another, transitions become rare. 
Typically, we look at the logarithm of this probability, which is called the 
\textit{free energy surface} (FES) or free energy profile as a function of CV $\mathbf{s}$:
\begin{equation}\label{eq:fes}
    F(\mathbf{s}) = -\frac{1}{\beta} \log p(\mathbf{s})
\end{equation}
We see that the free energy minima correspond to the metastable states, and the free energy barriers correspond to the transition states between them (Fig.~\ref{fig:probability-potential}A). 

Furthermore, the relative stability between two metastable states, A and B, can be measured by the free energy difference
\begin{equation}
    \Delta F = F_B - F_A = \frac{1}{\beta} \ln \frac{p_A(\mathbf{s})}{p_B(\mathbf{s})}=  \frac{1}{\beta} \ln \left ( \frac{\int_A d\mathbf{s} \: e^{-\beta F(\mathbf{s})}}{\int_B d\mathbf{s} \: e^{-\beta F(\mathbf{s})}} \right)
\end{equation}
where the integration domains correspond to the regions in CV space assigned to each metastable state (green and yellow highlighted regions in Fig.~\ref{fig:probability-potential}A and B). Note that the difference in free energy $\Delta F$ is hardly affected by the choice of CV, provided it can distinguish between the two minima~\cite{dietschreit2022obtain}. However, more care should be taken to extract the free energy barrier from FES, as the profiles depend on the choice of CV~\cite{dietschreit2022free}.

\paragraphtitle{Enhanced sampling to alter the probability.} 
To overcome the rare event problem, many enhanced sampling methods have been developed~\cite{Henin2022enhanced} with the idea of increasing the probability of these rare configurations.
Many of these methods rely on the identification of a small set of relevant CVs. 
Then, an external potential that depends on the atomic coordinates through the CVs, i.e., $V(\textbf{x}) = V(\textbf{s}(\textbf{x}))$, is added to the system to decrease the free energy barriers along the CVs and increase their fluctuations~\cite{valsson2016enhancing} (Fig.~\ref{fig:probability-potential}C).
Well-known examples of algorithms applying this principle are Umbrella Sampling~\cite{torrie1977nonphysical} and Metadynamics~\cite{laio2002escaping}.
From the probability perspective, the addition of the bias potential determines a new probability distribution  
which is sampled during the simulation, which is called \textit{biased distribution} (Fig.~\ref{fig:probability-potential}D):
\begin{equation}
    p_V(\textbf{x}) = \frac{e^{-\beta(U(\textbf{x})+V(\textbf{s}(\textbf{x})))}}{Z_V} \quad \text{with} \quad Z_V = \int \dd\textbf{x}\ e^{-\beta(U(\textbf{x})+V(\textbf{s}(\textbf{x})))}
\end{equation}
This is analogous to \textit{importance sampling} in statistics, which aims to estimate properties of a distribution by drawing samples from another one, typically easier to sample or more tractable.

\paragraphtitle{Reweighting.} 
A crucial aspect of the enhanced sampling procedure is the fact that we must be able to compute the properties of interest for the original system (such as its free energy surface) from the samples extracted from the modified distribution $p_V$. 
To do this, we need to apply a correction, or \textit{reweighting}. 
It is useful to derive this procedure in the case where the potential $V$ is static (or can be approximated as such) since this is an important ingredient of OPES.
For a generic quantity $O(\textbf{x})$, we would like to compute the ensemble average $\langle O(\textbf{x}) \rangle$ under the original Boltzmann distribution:
\begin{equation}
    \label{eq:expectation-value}
    \langle O(\textbf{x}) \rangle
        = 
        \int \dd \textbf{x}\ O(\textbf{x}) p(\textbf{x})
\end{equation}
using samples that are distributed according to $p_V(\textbf{x})$. 
We then need to express Eq.~\ref{eq:expectation-value} in terms of averages over the modified distribution. 
If we multiply the argument of the integral for $\frac{p_V(\textbf{x})}{p_V(\textbf{x})}$, after some calculations we find:
 \begin{equation}
        \langle O(\textbf{x}) \rangle
        = 
        \int \dd\textbf{x}\ \left( O(\textbf{x}) \frac{p(\textbf{x})}{p_V(\textbf{x})}\right) p_V(\textbf{x})
        = \frac{\langle O(\textbf{x}) e^{\beta V(\textbf{s}(\textbf{x}) ) } \rangle_V}{\langle e^{\beta V(\textbf{s}(\textbf{x}) ) } \rangle_V}
        \label{eq:reweighting}
    \end{equation}
where $\langle \cdot \rangle_V = \int \dd\textbf{x}\ \cdot p_V(\textbf{x})$ denotes the average under the biased distribution. However, while Eq.~\ref{eq:reweighting} connects the ensemble averages over the two different probability distributions, we need to translate this in terms of the samples we get from the biased MD simulations. 
The connection is made through the ergodic theorem~\cite{frenkel2001understanding}, which states that, for a sufficiently long time, the time average $\Bar{O}(\textbf{x})=\frac{1}{n_k}\sum^{n_k}_k O(\textbf{x}_k)$ of the properties of a dynamical system converges to their ensemble average $\langle O(\textbf{x}) \rangle$. 
Hence, we can rewrite Eq.~\ref{eq:reweighting} as a weighted time average:
    \begin{equation}
        \langle O(\textbf{x}) \rangle
        \approx \sum^{n_k}_k O(\textbf{x}_k) w(\textbf{x}_k)\quad\text{with}\quad w(\textbf{x}_k) = \frac{e^{\beta V(\textbf{s}(\textbf{x}_k))}}{\sum^{n_k}_k e^{\beta V(\textbf{s}(\textbf{x}_k))}}
        \label{eq:reweighting-samples}
    \end{equation}
where the samples $\textbf{x}_k$ are those obtained in an MD simulation performed with a bias potential $V$, and $w(\textbf{x}_k)$ are the normalized weights of the $k$-th configuration of the trajectory. 
The reweighting procedure then corresponds to correcting the weight of each sample to match that of the original distribution.
Indeed, in the limit of a large number of samples, the following relationship holds:
\begin{equation}
    w(\textbf{x}_k)\propto\frac{p(\textbf{x}_k)}{p_V(\textbf{x}_k)}
    \label{eq:reweighting-ratio}
\end{equation}

In particular, we can use Eq.~\ref{eq:reweighting-samples} to estimate the unbiased probability distribution $p(\textbf{s})$, and consequently the free energy profile (Eq.~\ref{eq:fes}). 
To do this, we note that the definition of the marginal of the density along $\mathbf{s}$ (Eq.~\ref{eq:density-cv}) can be rewritten as the ensemble average of a delta function, i.e., $p(\textbf{s})= \langle \delta(\textbf{s}-\textbf{s}(\textbf{x})) \rangle$.
Therefore, we obtain the following expression for the unbiased probability: 
    \begin{equation}
        p(\textbf{s}) \approx \sum^{n_k}_k \delta(\textbf{s}-\textbf{s}(\textbf{x}_k)) w(\textbf{x}_k)
        \label{eq:reweighting-probability}
    \end{equation}
From a practical point of view, the delta function can be estimated through a histogram, in which each sample is weighted by $w(\textbf{x}_k)$. Finally, we conclude with two remarks:
\begin{itemize}
    \itemsep0em
    \item Eq.~\ref{eq:reweighting-ratio} is general and can be applied to the reweighting between any two probability distributions. For example, these could be the Boltzmann distribution at two different temperatures $T_1$ and $T_2$, and in that case, it becomes $w(\mathbf{x})\propto e^{(\beta_2-\beta_1)U(\mathbf{x})}$. 
    \item The \textit{effective sample size} (ESS) can be used to measure the quality and efficiency of the reweighting procedure. The ESS quantifies the number of biased samples that contribute to the estimate of the unbiased quantity and can be calculated using the formula\cite{kong1994sequential,invernizzi2020unified}:
    \begin{equation}
        n_\mathrm{eff}=\frac{ 1 }{\sum^{n_k}_k w(\textbf{x}_k)^2}
        \label{eq:ess}
    \end{equation}
    In the ideal case, the ESS should scale proportionally to the number of samples of the trajectory, i.e., $n_\mathrm{eff}\propto n_k$. 
    On the other hand, a systematically low ESS (e.g., around 10) suggests that the enhanced sampling process is inefficient because only a few samples are contributing significantly to the estimate, i.e., there is not much density overlap between the biased $p_V(\textbf{x})$ and the original distribution $p(\textbf{x})$.
\end{itemize}

\subsection{OPES overview}

\begin{SCfigure}
        \centering
        \includegraphics[width=0.75\linewidth]{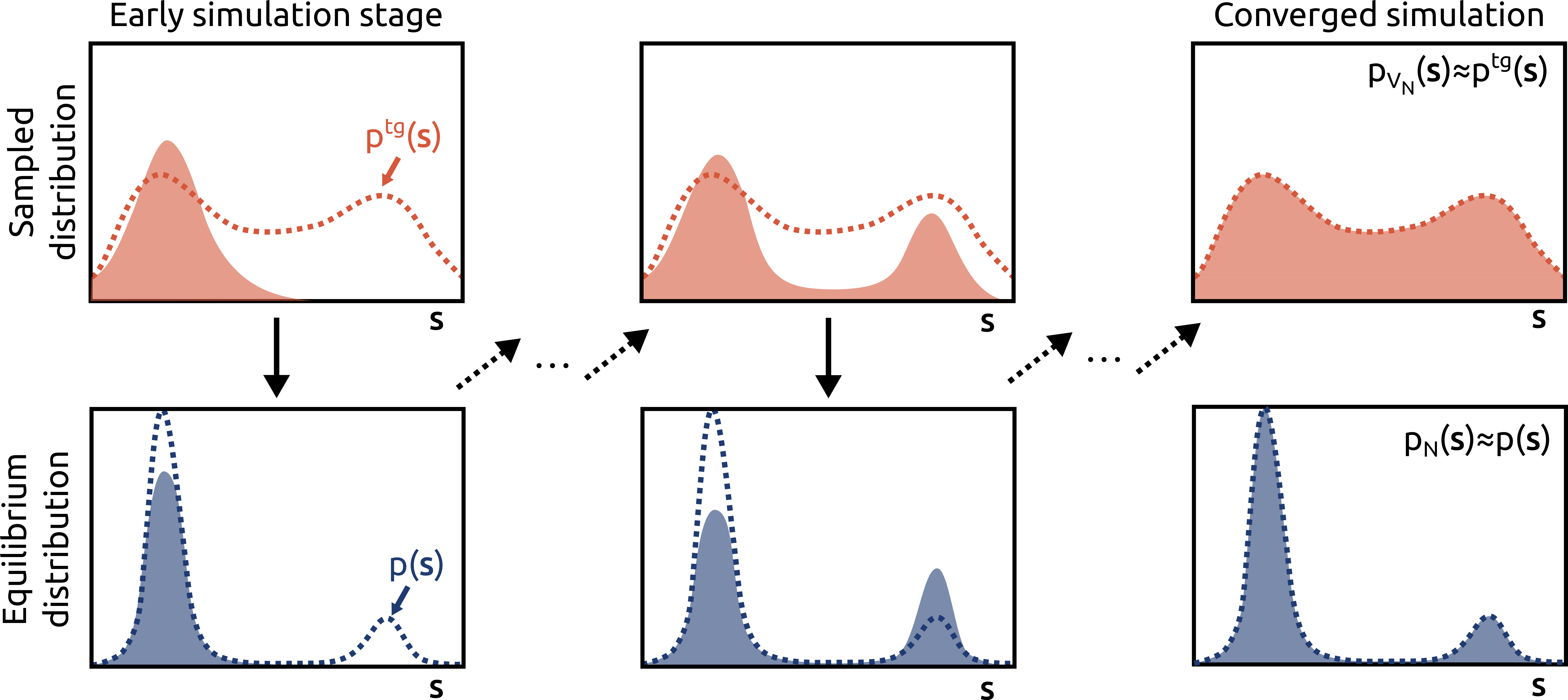}
        \caption{Schematic representation of the adaptive nature of OPES and its goals. At convergence, the sampled distribution corresponds to the target one, and the probability estimate corresponds to the unbiased equilibrium distribution.}
        \label{fig:OPES_iterations}
    \end{SCfigure}

\paragraphtitle{The general idea.}
The On-the-fly Probability Enhanced Sampling (OPES) method was designed with two goals in mind: to accelerate rare events and to do so in a way that makes it easy to retrieve unbiased statistical information about the system. 
To achieve them, OPES aims to sample a given target distribution $p^{tg}(\mathbf{x})$, in which the events of interest are no longer rare, but such that it still retains a good overlap with the original distribution $p(\mathbf{x})$ in configuration space.
This way, one can efficiently perform reweighting and obtain statistical information about $p(\mathbf{x})$.
This is realized by adding to the potential energy a bias $V(\mathbf{x})$ that in the limit is defined as:
\begin{empheq}[box=\fcolorbox{green!40!black!60!}{black!10}]{equation}
            \text{\texttt{OPES}:}\qquad V(\textbf{x}) = \frac{1}{\beta}\log \frac{p(\textbf{x})}{p^{tg}(\textbf{x})}
            \label{eq:opes_bias_general})
    \end{empheq}            
Since the distribution $p(\mathbf{x})$ is unknown at the beginning of the simulation, the bias is optimized in a self-consistent way, based on the following iterative procedure (see Fig.~\ref{fig:OPES_iterations}):
\begin{enumerate}
    \itemsep0em
    \item updating the estimates of the probability distribution $p(\textbf{x})$ via reweighting
    \item updating the bias potential $V(\textbf{x})$ accordingly
    \item sampling from the modified distribution with the potential $V(\textbf{x})$
\end{enumerate}

Equation~\ref{eq:opes_bias_general} describes a general framework, in which the main quantities are the estimate of the unbiased probability $p(\textbf{x})$ and the definition of the target distribution $p^{tg}(\textbf{x})$. 
Different realizations of these two aspects result in rather different behaviors, as we shall see in more detail below, discussing the variants of the method (see Sec.~\ref{sec:opes_variants}). However, before entering into the details of each variant, we highlight some important features of OPES.

\paragraphtitle{General features of OPES}: 
\begin{itemize}
\itemsep0em
    \item \textit{On-the-fly updates}. As in Metadynamics, OPES constantly updates the bias as new samples are gathered. This is more efficient than having to choose an observation time and then make a large change to the bias. 
    \item \textit{Quick convergence}. The iterative scheme aims at quickly reaching a regime of quasi-static bias. This way, most of the MD time can be used for statistical analysis. This might not be the case for Metadynamics, depending on the specific choice of parameters.
    \item \textit{Flexible target}. Depending on the variant, the target can be constructed in CV space with a predefined function (e.g., uniform distribution) or adjusted on the fly to broaden the equilibrium one (e.g., the well-tempered distribution). Alternatively, it can be used to target generalized ensemble distributions corresponding to replica exchange or umbrella sampling.
    \item \textit{Custom upper barrier}. OPES allows to set a limit on the FES barriers one wishes to overcome. This leads to a more precise definition of the target, focusing on the scale of free energies of interest and limiting exploration of unphysical states, by avoiding unnecessarily high bias.
    \item \textit{Replicas flexibility}. Contrary to methods like replica exchange, OPES does not require a minimum number of replicas to run in parallel. One can use any number of parallel simulations (multiple walkers) to concurrently optimize the same bias potential or opt for running several independent single OPES simulations to better estimate uncertainties.
    \item \textit{Ease of use}. The OPES method requires the scientist to set only a few physically relevant parameters since most internal parameters have robust default values or can be set automatically.
\end{itemize}

\subsection{OPES Variants}
\label{sec:opes_variants}
Thanks to its flexibility, several variants of OPES have been developed from the original formulation, which are useful for addressing different scenarios.
We want to provide a brief summary here to orient the reader before moving on to a detailed description. 
The purpose of this section is to make it clear in which scenario each variant should be used and what quantities they can give access to.  
These topics are also covered in the OPES masterclass \linkbox{TUTORIAL:\href{https://www.plumed-tutorials.org/lessons/22/003/data/NAVIGATION.html}{22.03}}, where you can experiment with the various parameters and compare the different variants.

    \begin{itemize}
        \itemsep0em 
        \item \sansbf{OPES-Metad~\cite{invernizzi2020rethinking}}. The original OPES formulation, which targets the well-tempered distribution and estimates the unbiased probability through reweighting. 
        This results in a bias that quickly becomes quasi-static, which is useful for converging free energies but requires nondegenerate CVs and an estimate of the barriers to be overcome.
        \item \sansbf{OPES-Explore~\cite{invernizzi2022exploration}}. The objective is the same as OPES-Metad, but the estimation of $p(\mathbf{x})$ is based on the sampled distribution and not on the reweighted one.
        This yields a bias that changes more over time, able to promote more transitions and explore the CV space even when the CVs are suboptimal. It scales better to higher dimensions and is more robust to the choice of the barrier and to degenerate CVs.
        It is useful for obtaining a quick, though not necessarily accurate, estimate of the free energy. It is the best choice for those familiar with metadynamics.
        \item \sansbf{OPES-Expanded~\cite{invernizzi2020unified}}. The target distribution is expressed as the sum of a series of overlapping probability distributions.
        This can be used to sample in one simulation the target distribution of replica exchange~\cite{sugita1999replicaexchange} corresponding to multiple temperatures, multiple pressures, or both. At the same time, it can be used to sample the target distribution of multiple windows umbrella sampling.
        \item \sansbf{OPES-Flooding~\cite{ray2022rare}}. The bias is built with the same scheme as OPES-Metad but, following the idea of conformational flooding, its action is limited to the reactant region. 
        This allows to harvest unbiased reactive trajectories and also estimate reaction rates.
    \end{itemize}

The next subsections are organized as follows. For each method, we first give a description and highlight the main quantities. We then describe the main parameters that need to be set in PLUMED, followed by practical suggestions and references to examples of how these methods have been used in the literature, pointing in particular to those that are present on PLUMED-NEST. 

\subsubsection{OPES-Metad: converging the free energy surface}
    \label{subsec:OPES-metad}
    \mybox{\paragraphtitle{When to use OPES-METAD?} We have a good understanding of the system (collective variables and free energy barrier) and want to obtain an accurate estimate of the free energy surface.\vspace{0.5em}}

OPES-Metad~\cite{invernizzi2020rethinking} is the first formulation of OPES, which is available in PLUMED with the command \texttt{OPES\_METAD}. Here, the target $p^{tg}$ is chosen as the well-tempered distribution in CV space $p^{WT}(\textbf{s})$~\cite{barducci2008welltempered} that aims to broaden the equilibrium probability $p(\textbf{s})$: 
    \begin{equation}
        p^{WT}(\textbf{s}) \propto \qty[p(\textbf{s})]^{1/\gamma}
    \label{eq:well-tempered}
    \end{equation}
where the parameter $\gamma$ (called bias factor) regulates the magnitude of the smoothing. 
In terms of free energy, the effective barriers in this distribution are decreased by a factor $\gamma$. 
The intuitive idea is to stay as close as possible to the original distribution, $p(\textbf{s})$, while reducing the barriers enough to obtain frequent transitions (see Fig.~\ref{fig:probability-potential} C and D). 
Compared to the uniform distribution ($\gamma=\infty$), it leads to a higher ESS.
This $p^{WT}(\textbf{s})$ is the same distribution that is sampled asymptotically in Well-Tempered Metadynamics~\cite{barducci2008welltempered}. 
        
In the OPES iterative scheme, the bias is optimized from an estimate of the unbiased probability $p(\textbf{s})$  which is obtained via \textit{reweigthing} (similar to Ref.~\citenum{mezei1987adaptive}). 
This is achieved using Eq.~\ref{eq:reweighting-samples} and employing a weighted kernel density estimation (KDE) to make $p(\textbf{s})$ continuous and differentiable. 
This means that, at step $n$, the probability is estimated via:
    \begin{equation}
        p_n(\textbf{s}) = \sum_k^{n_k} w_k G(\textbf{s}, \textbf{s}_k)
        \label{eq:opes_metad_kde}
    \end{equation}
where the normalized weights are computed from the bias $w_k \propto \exp\qty(\beta V_{k-1}(\textbf{s}_k))$ and $G(\textbf{s}, \textbf{s}_k)$ denote Gaussian kernels centered on the current position $\textbf{s}_k$.

From this estimate of the probability and following Eq.~\ref{eq:opes_bias_general}, the bias at step $n$ can be expressed as
    \begin{empheq}[box=\fcolorbox{green!40!black!60!}{black!10}]{equation}
            \text{\texttt{OPES\_METAD}:}\qquad V_n(\textbf{s}) = \left(1 - \frac{1}{\gamma}\right)\frac{1}{\beta}\ln\qty(\frac{p_n(\textbf{s})}{Z_n} + \epsilon
            \label{eq:opes_metad_bias})
    \end{empheq}            
in which $\epsilon\ll1$ is a regularization term to prevent the argument of the logarithm from going to zero, and the normalization factor $Z_n$ depends on the region of the CV space $\Omega_n$ that has been explored up to step $n$:
    \begin{equation}
        Z_n = \frac{1}{|\Omega_n|} \int_{\Omega_n} \dd\textbf{s}\ p_n(\textbf{s}) 
    \end{equation}
While this algorithm is conceptually simple, there are several details that are required to make it robust and efficient:
\begin{itemize}
    \itemsep0em
    \item \textit{Density estimation}. For the estimation of the probabilities, OPES uses KDE, which is a well-established density estimation method with only one important hyperparameter, the kernel bandwidth. 
    This approach is much better than simple histograms~\cite{mezei1987adaptive} and scales better than the Metadynamics approach of estimating the bias when multidimensional CVs are used.
    \item \textit{Kernel merging algorithm}. To keep the computational cost low, OPES uses a compression algorithm adapted from Ref.~\citenum{sodkomkham2016kernel}, which crucially prevents the number of kernels from growing linearly with the simulation time. 
    Contrary to a grid approach, it also scales better to higher dimensions by keeping kernels only in relevant regions of the CV space. 
    This also allows the bandwidth to shrink as the effective sampled size increases~\cite{invernizzi2020rethinking}, which makes the resolution of the OPES bias evolve from coarse to finer~\cite{invernizzi2022exploration}.
    \item \textit{Regularization term $\epsilon$}. Directly using the KDE estimate as argument for the logarithm in Eq.~\ref{eq:opes_metad_bias}, would lead to an unstable iterative scheme, with a bias that goes to infinity in regions far away from the initially sampled CV values. The regularization term $\epsilon$ not only fixes this issue but also allows for the introduction of an overall limit to the bias potential, see the \texttt{BARRIER} parameter.
    \item \textit{Normalization over explored CV space}, $Z_n$. One of the main drawbacks of having a bias that quickly becomes quasi-static is that this hinders the ability to quickly escape newly found metastable states~\cite{fort2017self}. 
    For instance, transition-tempered metadynamics~\cite{dama2014transition} solves this issue by asking the user beforehand what is the location in the CV space of the FES minima. 
    OPES handles this completely automatically via the $Z_n$ normalization factor by leveraging the fact that finding a new minimum (if the CVs are good) corresponds to increasing the explored volume.
\end{itemize}

    \paragraphtitle{Main parameters}:
    \begin{itemize}
        \itemsep0em
        \item \verb|PACE|: update stride of the bias. It should be high enough to allow the system to equilibrate with the instantaneous effective potential before the next update. For biomolecules simulations, it is often set \texttt{PACE}=500 MD steps.
        \item \verb|BARRIER|: expected height of the highest free barrier to cross in energy units. OPES sets an upper bound for the magnitude of the bias potential slightly higher than \texttt{BARRIER}. It is used to set $\gamma$ and $\epsilon$ to reasonable defaults.
        \item \verb|SIGMA|:  the initial width of the deposited kernels. It should be comparable with the standard deviation of the narrower basin in the CV space.
        If not specified, it is estimated adaptively by the OPES algorithm in the initial 10*\texttt{PACE} steps. Be careful when using a short \texttt{PACE}.
        \item \verb|NLIST|: use a neighbor list over the compressed kernels. This can significantly speed up the simulation with multidimensional CVs, so it is recommended to use it.
    \end{itemize}
    \paragraphtitle{Tips and tricks}:
    \begin{itemize}
        \itemsep0em
        \item \textit{CV quality}. Because the bias quickly reaches a quasi-static behavior, OPES requires good-quality CVs; otherwise, the simulation gets stuck in one of the metastable basins. Hence, OPES-Metad can be effectively used to detect suboptimal CVs, which can be improved, for example, using the framework described in Sec.~\ref{sec:mlcvs}. When that is not an option, it is more convenient to use OPES-Explore, which converges more slowly but is more robust to CV quality (see section~\ref{sec:opes_explore}). OPES-Explore can also be first used to get an estimate of free energy barriers to use as input for OPES-Metad. Or again, if there are small free energy barriers along other degrees of freedom, the use of the multithermal ensemble can be useful to take advantage of the higher ergodicity that exists at higher temperatures to overcome these barriers (see section~\ref{sec:opes_multithermal}). These techniques have also been combined in a replica exchange scheme called OneOPES to alleviate the issue of suboptimal CVs~\cite{rizzi2023oneopes}.

        \item \textit{Starting from the deepest basin}. We recommend starting the simulation from a configuration that corresponds to the deepest basin in the free energy landscape (if it is known) to better initialize the probability estimate. This usually makes convergence faster. 
        \item \textit{Running multiple independent replicas}. This is useful not only to estimate the uncertainty of the results but also to avoid the problem of unlucky initialization of the algorithm. 
        In fact, in the OPES-Metad scheme, an initial coarsen bias is determined quickly, followed by fine refinement. While this first guess is generally reliable, sometimes the algorithm can get it wrong, especially if the CV is poor. The best strategy to mitigate this potential problem is to run multiple simulations with the same PLUMED input but with different initializations in the MD code.
        \item \textit{Avoid large bias changes}. Due to the normalization $Z_n$, the OPES algorithm leads to an immediate deposition of a bias approximately similar to the value of the \verb|BARRIER| in the first steps. This behavior can be changed by not normalizing over the explored CV space, i.e., $Z_n=1$ in Eq.~\ref{eq:opes_metad_bias}, with the keyword \verb|NO_ZED|, which leads to a more incremental deposition of the bias. This could be useful, for instance, when using machine learning potentials at the very first stages when their extrapolation is not robust.
        \item \textit{Fixing sigma}. In the case of very long simulations, the kernel bandwidth might become excessively small, ultimately hitting on performance. In this case, it is suggested to use \texttt{SIGMA\_MIN} or \texttt{FIXED\_SIGMA}, which can be added during a restart.
    \end{itemize}

    \paragraphtitle{Examples}: See this tutorial for a quick start~\footnote{\url{https://www.plumed.org/doc-v2.9/user-doc/html/opes-metad.html}}.  Then, you can search for \texttt{OPES\_METAD} in the \linkbox{\href{https://www.plumed-nest.org/}{PLUMED-NEST}} repository and you will find many examples from the literature (together with the corresponding inputs).


    \subsubsection{OPES-Explore: exploring the landscape\label{sec:opes_explore}}
    \mybox{\paragraphtitle{When to use OPES-Explore?} When dealing with a novel system or CVs. We also suggest employing OPES-Explore when only suboptimal CVs are available (such CVs could later be improved using the framework described in Section \ref{sec:mlcvs})\vspace{0.5em}}
    
        As the name suggests, the \textit{Explore} variant, available in PLUMED with the command \verb|OPES_METAD_EXPLORE|, is geared toward faster exploration of the CV space. Furthermore, it still allows for good sampling when used with suboptimal CVs at variance with OPES-Metad (see Sec.~\ref{subsec:OPES-metad}). OPES-Explore is the OPES variant that will feel more familiar to anyone used to Metadynamics.
        
        In OPES-Explore, the target distribution is again the well-tempered one, but the unbiased probability $p(\textbf{s})$ is not estimated via reweighting, but rather from the biased one leveraging the relationship: $p(\textbf{s})\propto \qty[p^{WT}(\textbf{s}) ]^{\gamma}$. 
        The well-tempered distribution is approximated with the biased one $p_V(\textbf{s})$, and it is represented with a kernel density estimate (without reweighting), which at time $n$ this reads:
        \begin{equation}
            p^{WT}_n(\textbf{s}) = \frac{1}{n}\sum_k^n G(\textbf{s}, \textbf{s}_k)
        \end{equation}
        where again $G(\textbf{s}, \textbf{s}_k)$ are Gaussian kernels centered on the CV value $\textbf{s}_k$ for each step $k$. 
        Putting together all these ingredients, we obtain that the bias at iteration $n$ in OPES-Explore is given by:
            \begin{empheq}[box=\fcolorbox{green!40!black!60!}{black!10}]{equation}
                \text{\texttt{OPES\_METAD\_EXPLORE}: }\qquad V_n(\textbf{s}) = \qty(\gamma - 1) \frac{1}{\beta}\ln\qty(\frac{p^{WT}_n(\textbf{s})}{Z_n} + \epsilon)
            \end{empheq}
        and similarly to OPES-Metad, it still includes the bias factor $\gamma$, the regularization $\epsilon \ll 1$, and the normalization term $Z_n$ that depends on the extension of the sampling in the CV space.
        Tn OPES-Explore it is not possible to set $\gamma=\infty$ to sample a uniform distribution, as it is the case instead in OPES-Metad.
        
        \begin{figure}[h!]
                \centering
                \includegraphics[width=1\linewidth]{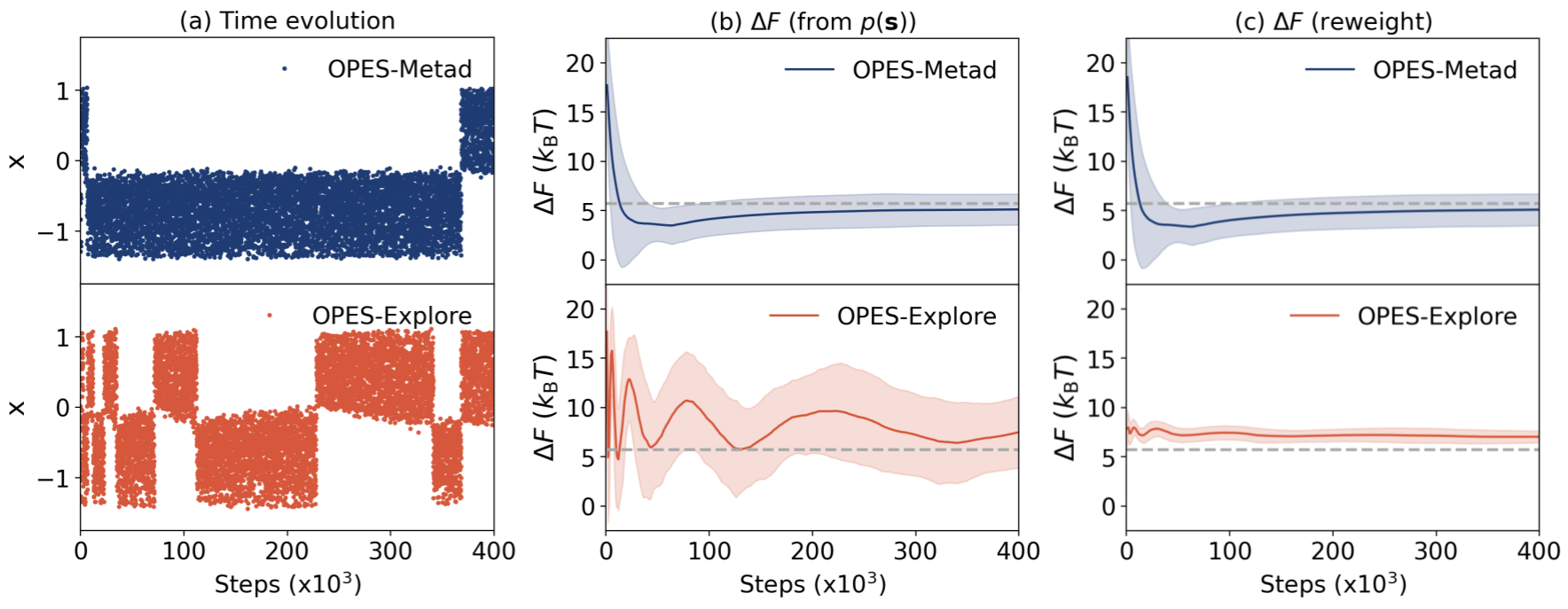}
                \caption{Comparison between OPES-Metad (top) and OPES-Explore (bottom) for a toy model system~ \cite{invernizzi2022exploration} using a suboptimal CV. We report (a) the time evolution of the biased CV and (b,c) the free energy difference $\Delta F$ between the two states. In (b) $\Delta F$ is calculated directly from the estimate of $p(\textbf{s})$, while in (c) via reweighting. The results are obtained from 25 independent replicas, showing the average (solid curve) and standard deviation (shaded area). Dashed lines indicate the reference value.   } 
                \label{fig:opes_explore}
        \end{figure}

    \paragraphtitle{Main parameters}:
    \begin{itemize}
    \itemsep0em
        \item \verb|BARRIER|: expected height of the free energy barrier to cross, in energy units. At variance with \texttt{OPES\_METAD}, it is less strict in respecting this \texttt{BARRIER}, and the magnitude of the bias potential can reach $\sim2\cdot$\verb|BARRIER|.
        \item \verb|SIGMA|: sets the initial width of the deposited kernels. For simplicity, it is designed to be chosen in the same way as for \texttt{OPES\_METAD}, looking at unbiased fluctuations of the CV.
        This value is then internally multiplied by $\sqrt{\gamma}$, to account for the fact that $p^{WT}$ has larger fluctuations. If absent, it is automatically estimated.
        \item See OPES-Metad main parameters.
    \end{itemize}
    \paragraphtitle{Tips and tricks}:
    \begin{itemize}
    \itemsep0em
        \item \textit{Faster high-dimensional exploration}. thanks to the reduced number of kernels used for constructing the probability, OPES-Explore is more suitable for the exploration of high-dimensional surfaces~\cite{invernizzi2022exploration}.
       \item \textit{Free energy estimates}. the FES can be obtained from OPES-Explore trajectories in two ways, either directly from the bias potential $
                F_n(\textbf{s}) = -(1-\frac{1}{\gamma})^{-1} V_n(\textbf{s})
            $ or via reweighting (Eq.~\ref{eq:fes}).
        At variance with the OPES-Metad case, in which these two approaches are equivalent, here they can differ significantly, especially in the first part of the simulation when the bias is far from convergence. Fig.~\ref{fig:opes_explore} shows the differences between the direct and reweighted estimates of the free energy difference in the case of the OPES-Metad and OPES-Explore variants.
        \item \textit{Driving suboptimal CVs}. Fig.~\ref{fig:opes_explore} also shows the different behavior of the two variants when a suboptimal variable is used. In the case of OPES-Metad, since the bias converges rapidly to a quasi-static regime, the system gets stuck into a local minima. 
        In the case of the OPES-Explore, the potential induces more transitions as a consequence of the bias changing over time, but this is not necessarily reflected in a better estimate of the free energy since the transitions occur further away from equilibrium. 
        At the same time, it still converges faster than Metadynamics (in general) and scales better with the dimensionality, thanks to the KDE estimator employed in OPES, which is more efficient than the bias estimator implemented in Metadynamics.
    \end{itemize}
    \paragraphtitle{Examples}:
    \begin{itemize}
    \itemsep0em
        \item \textit{Reaction discovery} \linkbox{NEST-\href{https://www.plumed-nest.org/eggs/22/004/}{22.004}}. OPES-Explore was used together with a variable derived from spectral graph theory to automatically find reaction products, running simulations with an increasingly higher \verb|BARRIER|~\cite{Raucci2022}. This was later applied also to molecular photoswitches~\cite{raucci2022photoswitches} \linkbox{NEST-\href{https://www.plumed-nest.org/eggs/22/038/}{22.038}} and to enzymatic catalyisis with QM/MM simulations~\cite{das2023enzyme} \linkbox{NEST-\href{https://www.plumed-nest.org/eggs/23/017/}{23.017}}. Typically, after the states are discovered, more tailored ML-based CVs are built and the free energy is converged with OPES-Metad.
    \end{itemize}

    \subsubsection{OPES-Expanded: generalized ensemble simulations }
    \label{sec:opes-expanded}
        In OPES-Expanded, available in PLUMED with the command \texttt{OPES\_EXPANDED}, the target distribution is not defined directly in the collective variable space but, more generally, is built as a combination of different ``versions'' of the original system, which might differ, for example, in the potential energy, temperature, pressure, or other quantities~\cite{invernizzi2020unified}.
        The ensembles generated by such target distributions are commonly known as expanded or generalized ensembles~\cite{lyubartsev1992new}.
        This idea can be formalized by considering a discrete set of values $\{\lambda\}$, which can be related to a single quantity (e.g., the temperature) or multiple ones, and define a set of corresponding single-ensemble probability distributions $p_\lambda(\textbf{x}) = e^{-\beta U_\lambda(\textbf{x})}/Z_\lambda$ where $Z_\lambda$ is the appropriate partition function.
        The distribution $p^{tg}_{\{\lambda\}}(\textbf{x})$ that we shall target with OPES-Expanded can be written as the sum over the single-ensemble distributions according to
        \begin{equation}
            p^{tg}_{\{\lambda\}}(\textbf{x}) = \frac{1}{N_\lambda}\sum_\lambda p_\lambda(\textbf{x})
            \label{eq:tg_prob_opes_expanded}
        \end{equation}
        In order to write the target distribution and bias potential in a general and compact fashion we will use the notion of \emph{expansion collective variables}, defined as $\Delta u_\lambda(\textbf{x}) = \beta \qty( U_\lambda(\textbf{x}) - U_0(\textbf{x}))$, where $U_0(\textbf{x})$ denotes the potential of the original system.
        Using the expansion collective variables, we can rewrite Eq.~\ref{eq:tg_prob_opes_expanded} as
        \begin{equation}
            p^{tg}_{\{\lambda\}}(\textbf{x}) = p_0(\textbf{x}) \frac{1}{N_\lambda}\sum_\lambda e^{-\Delta u_\lambda(\textbf{x}) +  \beta \Delta F(\lambda)}
            \label{eq:tg_prob_opes_expanded_expansion_cvs}
        \end{equation} 
        where $p_0(\textbf{x})=e^{-\beta U_0(\textbf{x})}/Z_0$ is the unbiased distribution, and $\Delta F(\lambda)=-\log(Z_\lambda/Z_0)$ is difference in free energy of the system $\lambda$ from the original one. 
        Using the definition of bias potential within OPES (Eq.~\ref{eq:opes_bias_general}), the OPES-Expanded bias potential at iteration $n$ can be expressed as:
        \begin{empheq}[box=\fcolorbox{green!40!black!60!}{black!10}]{equation}
                \text{\texttt{OPES\_EXPANDED}: }\qquad V_n(\textbf{x}) = -\frac{1}{\beta}\log\qty(\frac{1}{N_\lambda}\sum_\lambda e^{-\Delta u_\lambda(\textbf{x}) +  \beta \Delta F_n(\lambda)})
                \label{eq:opes_expanded}
        \end{empheq}
    where $\Delta F_n(\lambda)$ are the estimates of $\Delta F(\lambda)$ at iteration $n$ obtained through an on-the-fly reweighting of the sampled distribution (see Ref.~\citenum{invernizzi2020unified} for further details).

    Don't feel discouraged if the formalism above appears to be somewhat abstract. In practice, in PLUMED, the user only needs to define a small number of parameters.
    More in detail, one needs to specify one (or more) expansion CVs from those implemented, summarized in Table~\ref{tab:expanded_ensembles}, and pass them to the \texttt{OPES\_EXPANDED} function.
    In the following, to clarify the method and its practical usefulness, we will discuss, with the help of some simple examples, the multicanonical (i.e., multithermal) and multiumbrella target distributions, which are arguably the most widely used in practice.

    \paragraphtitle{Main general parameters}:
    \begin{itemize}
        \itemsep0em
        \item \verb|PACE|: update stride of the bias. It should be high enough to allow the system to equilibrate with the instantaneous effective potential before the next update, e.g., \texttt{PACE}=500 MD steps.
        \item \verb|OBSERVATION_STEPS|: number of unbiased initial \texttt{PACE} steps to collect statistics for the algorithm's initialization.  Often, during initialization, the $\{\lambda\}$ parameters are determined automatically based on a minimum and maximum value of $\lambda$ provided by the user. For this reason, don't cut corners! Always start OPES-Expanded simulations from a well-equilibrated configuration at the target temperature and pressure of the MD engine.
    \end{itemize}

        \setlength\tabcolsep{0pt}
        \renewcommand{\arraystretch}{1.2} 
        \begin{table}[h!]
        \caption{Summary of the ensembles that can be targeted with OPES-Expanded and their corresponding expansion CVs, parameters, and PLUMED command. Here $V(\textbf{x})$ denotes the volume, not the bias potential.}
        \label{tab:expanded_ensembles}
        \begin{tabular*}{\linewidth}{@{\extracolsep{\fill}} lcccc}
        \textbf{Target ensemble}         & \textbf{Expansion CVs}               & \textbf{Parameters} & \textbf{CVs}   & \textbf{PLUMED command}                \\
        \hline
        Linearly expanded       & $\lambda\Delta u(\textbf{x})$            & $\{\lambda\}$  & $\Delta u (\textbf{x})$ & \texttt{ECV\_LINEAR} \\
        Multicanonical          & $\qty(\beta - \beta_0)U(\textbf{x})$     & $\{\beta\}$    &   $U(\textbf{x})$        & \texttt{ECV\_MULTITHERMAL}  \\
        Multibaric              & $\beta_0\qty(p - p_0)V(\textbf{x})$      & $\{p\}$        &   $V(\textbf{x})$         & \texttt{ECV\_LINEAR} \\
        Multithermal-multibaric & $\qty(\beta - \beta_0)U(\textbf{x}) + \beta_0\qty(p - p_0)V(\textbf{x})$ &  $\{\beta, p\}$ & $U(\textbf{x}), V(\textbf{x})$ & \texttt{ECV\_MULTITHERMAL\_MULTIBARIC} \\
        Multiumbrella           & $\qty(s(\textbf{x}) - s_\lambda)^2/(2\sigma^2)$  & $\{s_\lambda\}$       &  $s(\textbf{x})$ & \texttt{ECV\_UMBRELLAS\_LINE} 
        \end{tabular*}
        \end{table}
    
    \paragraphtitle{\normalsize{A. Multithermal expansion}} 

    \vspace{0.5em}
    \mybox{\paragraphtitle{When to use OPES-Expanded with a MULTITHERMAL expansion?} There are mainly two use cases.  The first is to exploit the higher ergodicity of higher temperatures to drive transitions between metastable states. The second is to obtain sample configurations at different thermodynamic conditions in a single simulation. 
    \vspace{0.5em}}
        \label{sec:opes_multithermal}
        The goal of multicanonical simulations is to sample all relevant configurations of the system over a range of temperatures.
        A notable example is replica exchange\cite{sugita1999replicaexchange}, which performs and combines simulations of parallel replicas of the system at different temperatures.
        However, in many cases, this is not feasible because the number of replicas required is too large, which has motivated the development of methods in which a range of temperatures is sampled in a single simulation~\cite{okumura2004molecular,okumura2004monte,shell2002generalization,piaggi2019calculation,piaggi2019multithermal}.
        The OPES framework allows to easily construct a suitable bias potential to sample a multicanonical target distribution in a single simulation, via the OPES-Expanded scheme. 
        Afterward, the simulation can be reweighted to calculate static observables at any temperature within the targeted temperature range.
        Let's first review the underpinnings of this particular case of OPES-Expanded and then illustrate its workings with an example.
        
        We consider a system simulated at inverse temperature $\beta_0$, and we aim to sample configurations in a range of inverse temperatures $\beta$ such that $\beta_{min}<\beta<\beta_{max}$.
        The multithermal ensemble can be targeted with OPES-Expanded using the expansion CVs
            \begin{equation}
                \Delta u_{\beta_i}(\textbf{x}) = (\beta_i - \beta_0)U(\textbf{x})
                \label{eq:multithermal_expansion_cvs}
            \end{equation}
        which are available in PLUMED via the command \texttt{ECV\_MULTITHERMAL} and where $i=1,..,N_\beta$, and $\beta_{min}<\beta_i<\beta_{max}$ covers the desired range of temperatures.
        These expansion CVs are then used to compute the bias according to Eq.~\ref{eq:opes_expanded}.
        Note that the expansion CVs defined in Eq.~\ref{eq:multithermal_expansion_cvs}, and thus also the bias potential, depend on $\textbf{x}$ only through the potential energy $U(\textbf{x})$.
        The fact that the multithermal ensemble can be sampled using the energy as CV is a key insight put forward in Ref.~\citenum{piaggi2019multithermal}.
        Furthermore, using Eq.~\eqref{eq:tg_prob_opes_expanded_expansion_cvs} and with a little algebra, we can also see that these expansion CVs lead to the target distribution
        \begin{equation}
            p^{tg}(\textbf{x}) = \frac{1}{N_\beta}\sum_{i=1}^{N_\beta} \frac{e^{-\beta_i U(\textbf{x})}}{Z_{\beta_i}}.
        \end{equation}
        where $Z_{\beta_i}$ is the partition function at inverse temperature $\beta_i$.
        Thus, the target distribution is simply a sum of Boltzmann distributions at different temperatures $\beta_i$ within the chosen temperature range.
        
\begin{figure}[h!]\centering
        \includegraphics[width=0.85\columnwidth]{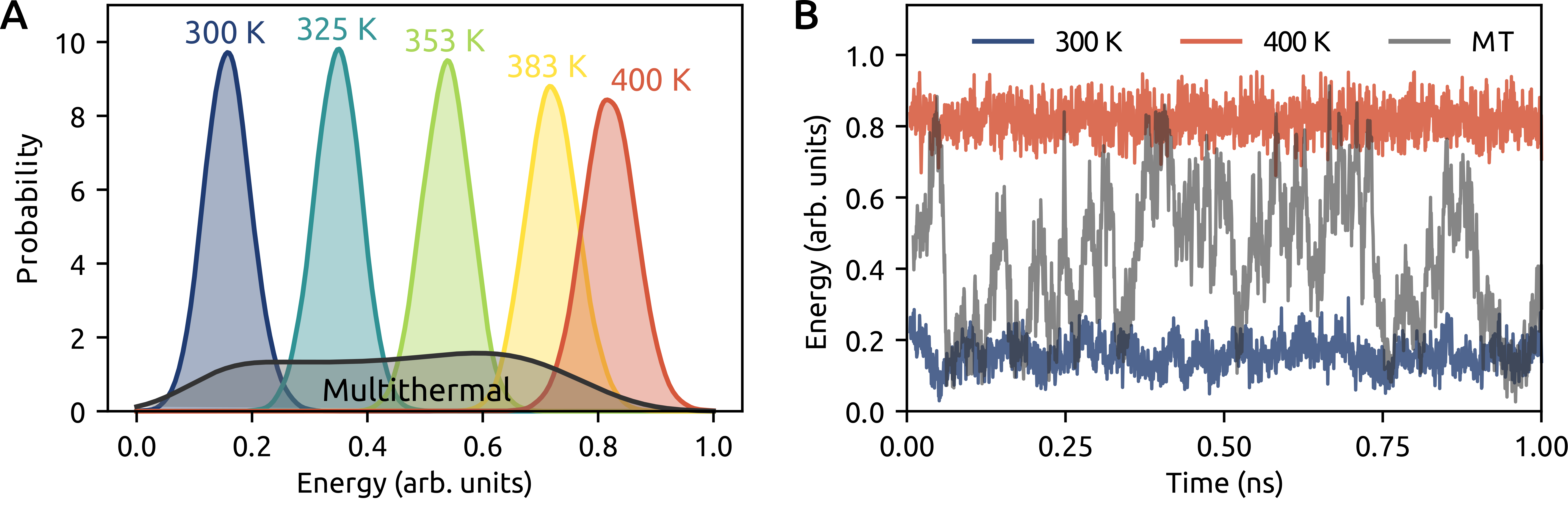}
        \caption{Example of OPES-Expanded with the multithermal expansion. \textbf{A}. shows the probability distributions of the energy for five simulations at constant temperatures in the range 300~K to 400~K together with  the distribution sampled by OPES-Expanded multithermal, which covers the whole range in a single simulation. \textbf{B}. depicts the energy as a function of simulation time in constant temperature simulations with target temperatures of 300~K and 400~K and in a multithermal one.}
        \label{fig:opes-mt}
        \end{figure}

        Let us now illustrate how the algorithm works in practice through a pedagogical example, using as a system a box with 288 water molecules in the liquid phase and a target temperature range of 300~K to 400~K (the simulation setup follows Ref.~\citenum{piaggi2021phase}).
        In Fig.~\ref{fig:opes-mt}A, we show the unbiased probability distributions of the potential energy at five different temperatures and the sampled multithermal distribution, which covers the entire range.
        We also show in Fig.~\ref{fig:opes-mt}B the potential energy as a function of time.
        In the multithermal (MT) simulation, the initial energy is characteristic of the system at the target temperature of the thermostat, i.e., 350~K.
        Initially, the bias potential quickly forces the system to explore energies compatible with the highest temperature (400~K) and then back to energies expected for the lowest temperature (300~K).
        For reference, we also show the results of simulations at constant temperatures of 300 K and 400~K, where we observe that the sampled energy range is much smaller.
        It can be seen that, after the initial transient, the algorithm explores reversibly the expected range of energies for temperatures between 300~K and 400~K. 

    \paragraphtitle{Specific parameters}:
    \begin{itemize}
        \itemsep0em
        \item \verb|TEMP_MIN| and \verb|TEMP_MAX| set the temperature range for the multithermal expansion. The number of steps, which are geometrically spaced, by default, is automatically determined based on the \texttt{OBSERVATION\_STEPS}~\cite{invernizzi2020unified}.
    \end{itemize}
    \paragraphtitle{Tips and tricks}:
    \begin{itemize}
        \itemsep0em
        \item The temperature $\beta_0$ set by the MD thermostat is not altered. The effect of the potential is to sample configurations that are relevant at other temperatures by biasing the potential energy while the temperature remains the same.
        \item To obtain the estimated free energy at temperature $\beta$ different from the simulated one $\beta_0$, a term that depends on the difference between the two temperatures must be added to the statistical reweighting: $w(\mathbf{x})\propto\exp(\beta V(\mathbf{x}) + (\beta_0-\beta) * U(\mathbf{x})) $.
        \item In the case of constant-pressure simulations, the potential energy must be replaced with the enthalpy. This can be achieved easily by adding the pressure-volume term $P\:V(\textbf{x})$ to the potential energy. 
        For further details, see the PLUMED manual entry for \verb|ECV_MULTITHERMAL|.
    \end{itemize}
    \paragraphtitle{Examples}:
    \begin{itemize}
        \itemsep0em
        \item \textit{RNA folding} \linkbox{NEST-\href{https://www.plumed-nest.org/eggs/22/043/}{22.043}}: OPES Multithermal has been shown to be more computationally efficient than parallel tempering metadynamics to study the folding of RNA tetraloops~\cite{rahimi2023comparison}.
        \item \textit{Chignolin folding with DeepTICA} \linkbox{NEST-\href{https://www.plumed-nest.org/eggs/21/039/}{21.039}}: Multithermal simulations can serve as collective variable-free exploration method which is used to extract relevant CVs (see Sec.~\ref{sec:mlcvs})~\cite{bonati2021deep}.
    \end{itemize}

    \vspace{0.5em}
    \paragraphtitle{\normalsize{B. Multiumbrella expansion}}.
    \vspace{0.5em}
    
    \mybox{\paragraphtitle{When to use OPES-Expanded with a MULTI-UMBRELLA expansion?} Useful to focus the sampling in a given region of the CV space without using restraints such as harmonic walls. Also, the multiumbrella expansion is useful to enhance the sampling along a CV while simultaneously using the multithermal expansion to exploit the higher ergodicity.\vspace{0.5em}}
    
        Umbrella sampling is a pioneering enhanced-sampling approach proposed by Torrie and Valleau, and it is still widely used today to reconstruct the free energy of a system along a given collective variable $\textbf{s}$\cite{torrie1977nonphysical}.
        This approach is based on running multiple simulations in which the system is confined within narrow regions of CV space by suitable harmonic bias potentials.
        The results of such simulations can then be combined using some post-processing analysis, such as the weighted histogram analysis method (WHAM)\cite{kumar1992weighted}.
        If one considers the collection of such simulations as an expanded ensemble, the same result can be obtained in a single simulation using the OPES-Expanded formalism, thus making the reweighting of the results straightforward.
        Furthermore, choosing the number and location of the umbrellas is easier, and having a single longer simulation can also help with sampling orthogonal slow degrees of freedom.

        To target the so-called multiumbrella expanded ensemble, suitable expansion CVs can be defined as
            \begin{equation}
                \Delta u_{\lambda} (\textbf{x}) = 
                \frac{\qty(\textbf{s}(\textbf{x}) - \textbf{s}_\lambda)^2}{2\sigma^2}
            \end{equation}
        for uniformly distributed and overlapping umbrellas (harmonic potentials) of width $\sigma$ and positions $\textbf{s}_\lambda$.
        In PLUMED, such an expansion CV can be defined using the \verb|ECV_UMBRELLAS_LINE| function.
        Note that the target distribution in this case is,
        \begin{equation}
            p^{tg}(\textbf{x}) = \frac{1}{N_\lambda}\sum_{i=1}^{N_\lambda} \frac{e^{-\beta U(\textbf{x}) - \qty(\textbf{s}(\textbf{x}) - \textbf{s}_\lambda)^2 / 2\sigma^2 }}{Z_{\lambda}},
        \end{equation}
        which is indeed the sum of the probability distributions sampled in each window of the umbrella sampling technique.

    \paragraphtitle{Specific parameters}:
    \begin{itemize}
    \itemsep0em
        \item \verb|CV_MIN| and \verb|CV_MAX| set the CV range that the multiumbrella expansion targets. Note that the ability to define minimum and maximum CV values to explore is useful in many cases where the sampling needs to be focused.
        \item \verb|SIGMA| controls the width of the Gaussians, which can be chosen based on the unbiased fluctuations of the CV. It can also be chosen manually to have more control over the target distribution. Large values of \verb|SIGMA| lead to a smooth and broad target distribution, which nonetheless preserves some features of the unbiased distribution. Instead, as \verb|SIGMA| is reduced, the distribution approaches a uniform distribution in CV space.
        \item \verb|SPACING| controls the distance between the Gaussians, in units of \verb|SIGMA|, by default equal to 1. A value of $1.5$ or $2$ can reduce the total number of umbrellas, often without any drawback.
    \end{itemize}
    
    \paragraphtitle{Tips and tricks}:
    \begin{itemize}
    \itemsep0em
        \item This can be combined with multithermal expansion to simultaneously enhance sampling based on the temperature and the CV.
        \item This technique is particularly suited to be used in combination with MLCVs because it allows enhancing the sampling only in the CV interval that is well represented in the training dataset instead of pushing the system outside such a region.
        \item It is possible to use \texttt{ECV\_MULTIUMBRELLAS\_FILE} to have complete control on the position and size of each of the umbrellas. This can be useful, for example, to define a path in high-dimensional CV space.
    \end{itemize}

\begin{figure}[h!]\centering
            \includegraphics[width=0.85\columnwidth]{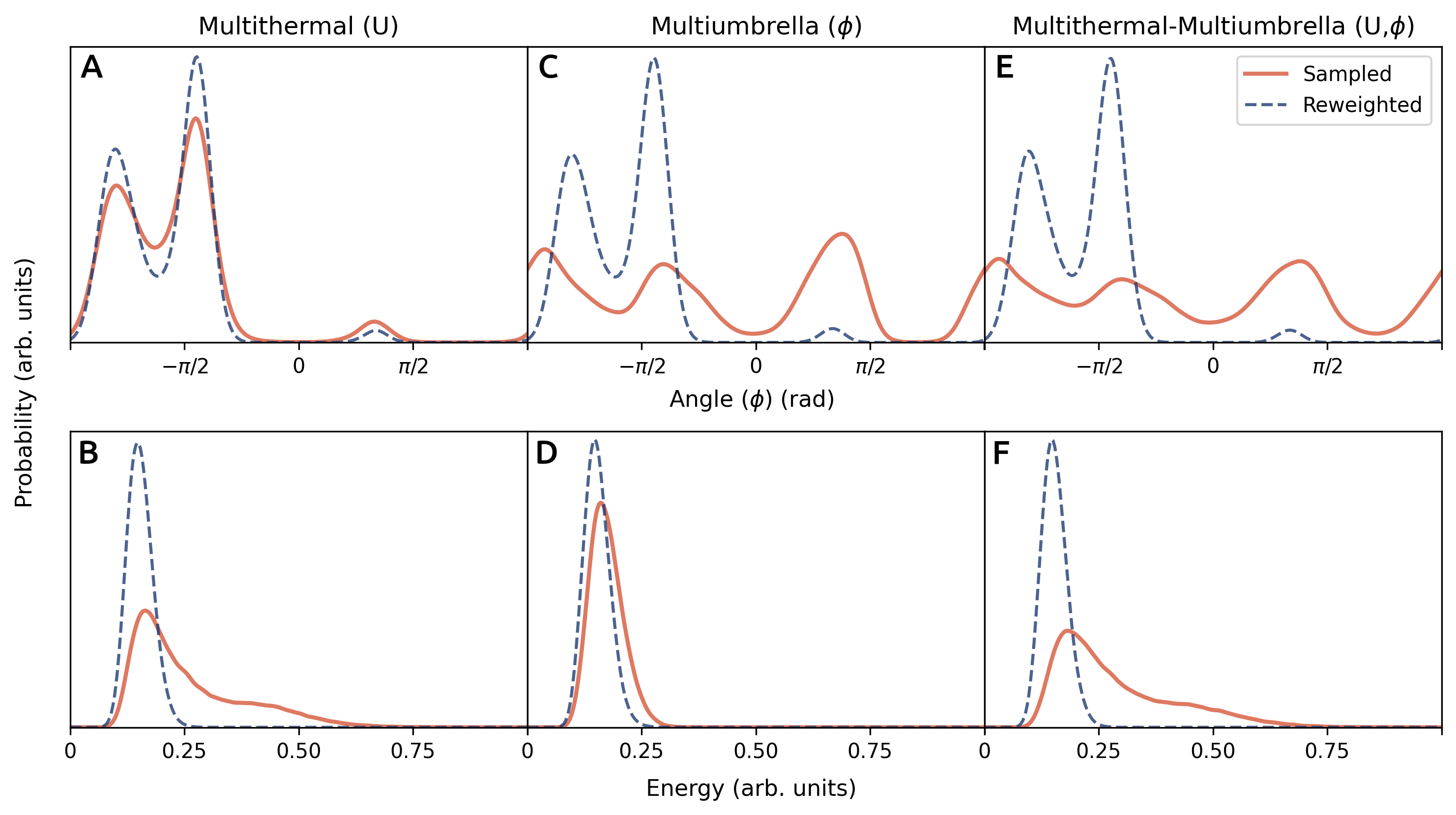}
            \caption{Example of the multithermal (left) and multiumbrella (center) ensembles, and their combination (right), generated by OPES-Expanded and applied to alanine dipeptide in vacuum. 
            In each case, we show the sampled and reweighted distributions of the torsion angle $\phi$ (above) and potential energy $U$ (below). See text for further details.}
            \label{fig:opes-target}
        \end{figure}

    \paragraphtitle{Comparison between Multithermal and Multiumbrella}.
    To better understand their differences, we use an example taken from Ref.~\citenum{invernizzi2021opes}, which considered the multithermal and multiumbrella expanded ensembles and also their combination in the case of the well-known conformational equilibrium of alanine dipeptide in vacuum\cite{invernizzi2021opes}. In particular, we can look at how the probability distributions of a relevant CV (the $\phi$ torsional angle, top row) and the energy (bottom row) are altered by the different biases. 
    Let's start first with the multithermal case.
        We show in Fig.~\ref{fig:opes-target} A and B, the sampled and reweighted (unbiased) distributions of $\phi$ and the potential energy $U$.
        Here, the sampled $U$ distribution is significantly broadened with respect to the reweighted one, while the $\phi$ distribution is only somewhat smoothed.
        The new (sampled) distribution will enhance the fluctuations of the CV $\phi$ and thus help to overcome the bottlenecks along that CV, but not in a very targeted or efficient fashion.
        The next case, the multiumbrella ensemble along $\phi$, is shown in Fig.~\ref{fig:opes-target}.
        Panels C and D show the $U$ and $\phi$ distributions, respectively.
        Within the multiumbrella ensemble, the $U$ distribution is almost unchanged, yet the bottlenecks along $\phi$ have been significantly reduced. 
        Finally, if the multiumbrella and multithermal ensembles are combined, the sampled distributions of both CVs ($U$ and $\phi$) are significantly broadened, allowing for a much better sampling (see Fig.~\ref{fig:opes-target} E and F).

    \subsubsection{OPES-Flooding: computing kinetic rates }
    \label{sec:opes_flooding}
    \mybox{\paragraphtitle{When to use OPES-FLOODING?} When you want to recover kinetic rates and/or collect a set of transition paths. It requires a CV that can distinguish the transition state from the initial one and an estimate of the free energy barrier.\vspace{0.5em}}
    
\begin{SCfigure}[][h!]
                \centering                \includegraphics[width=0.6\linewidth]{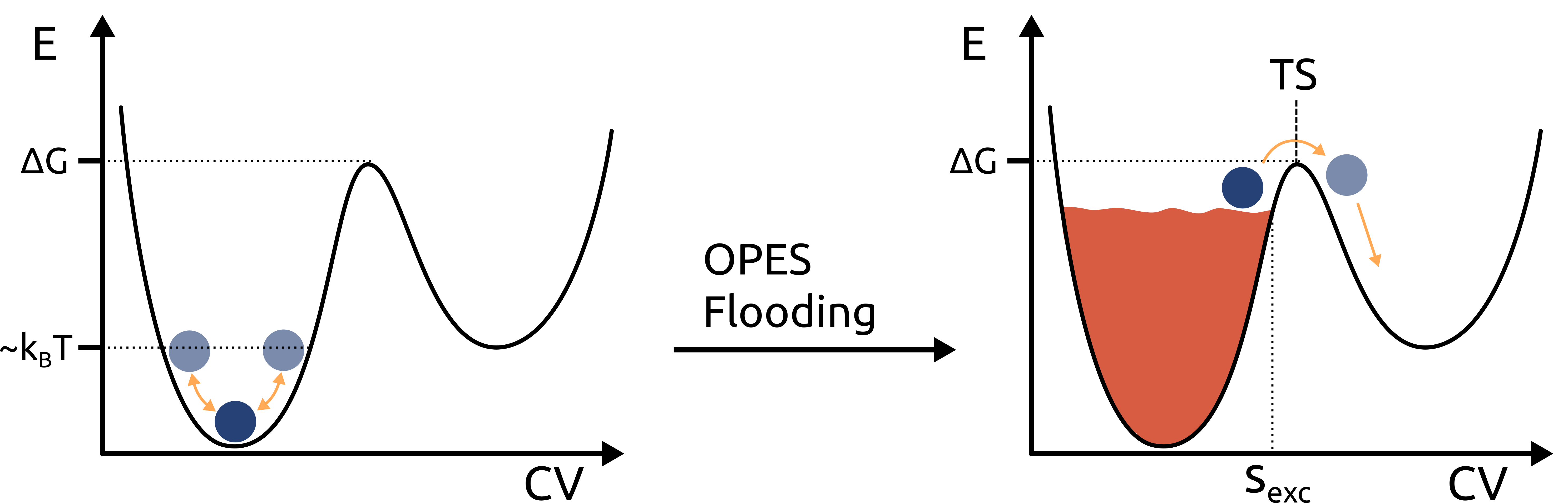}
                \caption{Visualization of the working principle of OPES-Flooding simulations. A bias potential is applied to partially fill the starting basin without affecting the transition state TS region.
                This way, unbiased reactive trajectories can be collected within an affordable computational time. }
                \label{fig:opes_flooding}
            \end{SCfigure}

        The OPES-Flooding~\cite{Ansari2022, ray2022rare} variant was developed for efficiently collecting unbiased transition and studying kinetic rates by combining the OPES biasing scheme and the ideas of conformational flooding~\cite{grubmuller1995flooding,mccarty2015variationally} and hyperdynamics~\cite{voter1997hyper}. In these works, the authors demonstrated that kinetic and dynamical information could still be recovered from biased simulations if the transition state (TS) region is left unaltered by the bias potential, i.e., no bias is applied there. This also motivated the infrequent metadynamics~\cite{tiwary2013metadynamics} method, in which small Gaussians are added with a very high stride to minimize the probability of interfering with the transition state.
        OPES-Flooding builds on this concept, realizing in a straightforward way the requirement of not perturbing the transition state due to its flexibility. 
        The OPES-Flooding idea, which we schematically depict in Fig.~\ref{fig:opes_flooding}, is to build a bias $V_f(\textbf{s})$ along a CV $\textbf{s}$ that is aimed at \emph{partially filling} only one of the metastable basins such that we can escape from it in a reasonable time and that the TS region is not affected by the bias itself.
        This last condition is ensured by the introduction of an additional parameter, the so-called \emph{excluded region} $\textbf{s}_{exc}$, and imposing that no bias is deposited for $\textbf{s} \in \textbf{s}_{exc}$. On the other hand, for $\textbf{s} \not\in \textbf{s}_{exc}$, the deposited bias still increases the probability of escaping from the minima and observing the transition.
        In practice, performing a set of simulations following this approach, one can collect a number of unbiased reactive trajectories that, besides providing TS-related configurations, can be used to statistically recover the unbiased kinetics of the process~\cite{ray2022rare, ray2023kinetics}.
        Indeed, if the above conditions are fulfilled, the ratio between the unbiased mean first passage time (MFPT) $t$ and the accelerated MFPT in the biased simulation $t_f$ is given by
            \begin{equation}
                \frac{t}{t_f} = \left\langle e^{\beta V_f(\textbf{s})} \right\rangle_{V_f}
            \end{equation}
        where the right-hand-side ensemble average is computed on the biased energy landscape.
        Such a quantity is referred to as the \textit{acceleration factor}, which provides a first measure of the computational efficiency of the adopted flooding protocol, i.e., the higher, the better.
        On the other hand, to evaluate the reliability of the estimated kinetics, one should first fit the results to the ideal Poisson distribution that $t$ should follow, which is $P(t) = \frac{1}{\tau}e^{-t/\tau}$, and compute the characteristic time of the transition $\tau$. Then, one should assess the fit quality using the standard Kolmogorov-Smirnov test and check the obtained \textit{p-value}~\cite{salvalaglio2014assessing}, which should be at least larger than 0.05 to be reliable.
        As the computation of kinetic rates is not the main focus of this Chapter, we refer the Reader to Refs.~\citenum{ray2022rare, ray2023kinetics} for further details on the topic.

    \paragraphtitle{Main parameters}.
    To use the OPES-Flooding method in PLUMED one should use the \verb|OPES_METAD| command with the additional following parameters:
    \begin{itemize}
    \itemsep0em
        \item \verb|EXCLUDED_REGION|: define the range of CV values $s_{exc}$ to be excluded to avoid applying bias on the transition state. This requires a function that is different from zero for $\textbf{s}$ where the bias should not be applied.        
        \item \verb|BARRIER|: in OPES-Flooding, at variance with the original OPES-Metad method, this parameter should be set to be lower than the actual free energy barrier of the system, to avoid depositing bias on the transition state region.
    \end{itemize}
    \paragraphtitle{Tips and tricks}:
    \begin{itemize}
    \itemsep0em
        \item \textit{Stopping the simulation}: as soon as the system reaches the final state, the simulation can be stopped by PLUMED using the \verb|COMMITTOR| keyword.
        \item \textit{Choice of the CV}: for flooding simulations, it is crucial that the CV can distinguish the starting metastable state from the transition state, but correctly describing the entire pathway may not be necessary.
        \item \textit{Kinetic estimates reliability}: one needs to find the best trade-off between acceleration and accuracy. Lower \verb|BARRIER| values and more conservative \verb|EXCLUDED_REGION| guarantee more accurate results but may require longer simulations. 
        The quality of the results should always be checked with the \textit{p-value} from the Kolmogorov-Smirnov test.
        \item \textit{Collecting reactive pathways}: even if OPES-Flooding was originally developed for kinetics calculations, it can also be used to generate unbiased reactive trajectories escaping from a known metastable state~\cite{ray2023deep,perego2024data}.
    \end{itemize}
    \paragraphtitle{Examples}:
    \begin{itemize}
    \itemsep0em
        \item \textit{Ligand-binding} \linkbox{NEST-\href{https://www.plumed-nest.org/eggs/22/017/}{22.17}}: OPES-Flooding was used to estimate the residence times of Benzamidine to Trypsin, also using machine learning-based CVs to capture the role of water in the process~\cite{Ansari2022}. 
        \item \textit{Chignolin folding} \linkbox{NEST-\href{https://www.plumed-nest.org/eggs/22/031/}{22.31}}: The folding and unfolding of the mini-protein chignolin has been studied using different ML-based CVs. Furthermore, OPES-Flooding has also been used as a way to collect transition state ensemble data to augment classifier-based CVs in the TPI-DeepTDA approach~\cite{ray2023deep} (see Sec.~\ref{sec:cvs_deepTDA}) \linkbox{NEST-\href{https://www.plumed-nest.org/eggs/23/009/}{23.009}}.
        \item \textit{Enzymatic catalysis} \linkbox{NEST-\href{https://www.plumed-nest.org/eggs/24/001/}{24.01}}: A QM/MM approach was used to calculate the kinetics of two prototypical enzymatic reactions: the transformation of chorismate to prephenate chorismate mutase, and the hydrolysis of maltopentaose by $\alpha$-amylase.~\cite{ray2024catalysis}  
    \end{itemize}

\subsection{Assessing convergence} 
From a theoretical point of view, convergence of an MD simulation happens when the sampled trajectory is ergodic and the time-average over it corresponds to an average over the phase space. In practical applications, this is never strictly true, and one typically has to assess if the convergence reached is good enough for its purposes. 
From a more practical point of view, below we suggest some quantities to monitor and/or strategies to follow to evaluate the convergence of OPES simulations.

\begin{itemize}
    \item The number of transitions between metastable states is an important quantity to monitor to assess convergence. However, it is important to understand that a time-varying bias can induce more transitions without leading to a better estimate of equilibrium probabilities (see fig.~\ref{fig:opes_explore}).  In fact, it is normal to observe more transitions in the early part where the bias is adjusting, than in the following one. Therefore, it is important to make sure that we get reversible transitions even after the initial transient (see below). To give a reference, we could say that the goal is to see a minimum of four reversible transitions after the transient. However, a larger number of transitions is recommended.
    \item Monitoring the time evolution of the estimate of relevant properties, such as the free energy difference between the basins of interest, is one of the best ways to assess convergence.
    \item Achieving quasi-static regime in OPES is signaled by the variable $c(t)$, corresponding to the ratio of the  partition functions for the biased and unbiased distributions, which must reach a constant value:
    \begin{equation}
        c(t) = -\frac{1}{\beta}\ln \frac{\int \dd\textbf{s}\  p(\textbf{s}) e^{-\beta V(\textbf{s},t)} }{\int \dd\textbf{s}\  p(\textbf{s})}  \approx \frac{1}{\beta} \log \langle e^{\beta V}\rangle      
    \end{equation}
    conveniently, $c(t)$ is always a scalar, thus can be easier to monitor than a high dimensional bias.
    \item If simulations made with multiple replicas are to be combined, either in the case of multiple walkers sharing the same bias or independent ones, it would be ideal for each to have reached convergence.
    \item To estimate the uncertainty on equilibrium properties (such as FES) from biased simulations a weighted block average should be performed, see Ref.~\citenum{bussi2019analyzing} and Appendix B of Ref.~\citenum{invernizzi2020unified}).
    \item The effective sample size (Eq.~\ref{eq:ess}) is a useful way to quantify the number of samples contributing to the reweighting and assess sampling efficiency. ESS is particularly useful for understanding whether multithermal simulations really provide samples over the full range of expected temperatures\cite{invernizzi2020unified,rahimi2023comparison,piaggi2021phase}.

\end{itemize}
\section{Machine learning collective variables}
\label{sec:mlcvs}
The other key ingredient of many enhanced sampling schemes are the collective variables (CVs). Indeed, without high-quality CVs, the enhanced sampling procedure will not be very effective. As we have shown in Sec.~\ref{sec:OPES}, some variants are more appropriate in the case where the variables are less optimal, however wherever possible, the goal should always be to improve them. In fact, the acceleration that can be achieved if we target all degrees of freedom that have relevant barriers is exponential.  It is, therefore, of paramount importance to understand how to identify CVs. For an overview of the different approaches and requirements, we suggest starting with Ref.~\citenum{bussi2020using}. In particular, in this section, we focus on data-driven approaches to learn CVs directly from the data, rather than having to guess them based on physical intuition. 

\subsection{Collective variables for enhanced sampling}
\label{sec:cvs_enhanced_sampling}

\paragraphtitle{CVs requirements}. 
As we introduced in~\ref{sec:free_energies}, we refer to collective variables as functions of atomic positions. In the context of enhanced sampling, we have two additional formal requirements: on the one hand, this function must be continuous and differentiable across configurational space. 
This is important because the external force due to the biasing potential $f=-\nabla _\mathbf{x}V(\mathbf{s}(\mathbf{x}))$ depends on both the derivatives of the bias w.r.t. the CVs $\mathbf{s}$ and the derivative of the CVs w.r.t. the atomic positions $\textbf{x}$.
On the other hand, the CV should also respect the relevant symmetries of the process under investigation, typically rotational and translational invariance and sometimes also the permutational one. 
This ensures that the CV correctly represents the physical properties of the system and that the biasing potential does not introduce artifacts into the simulation.

\begin{figure}[h!]\centering
    \includegraphics[width=1\columnwidth]{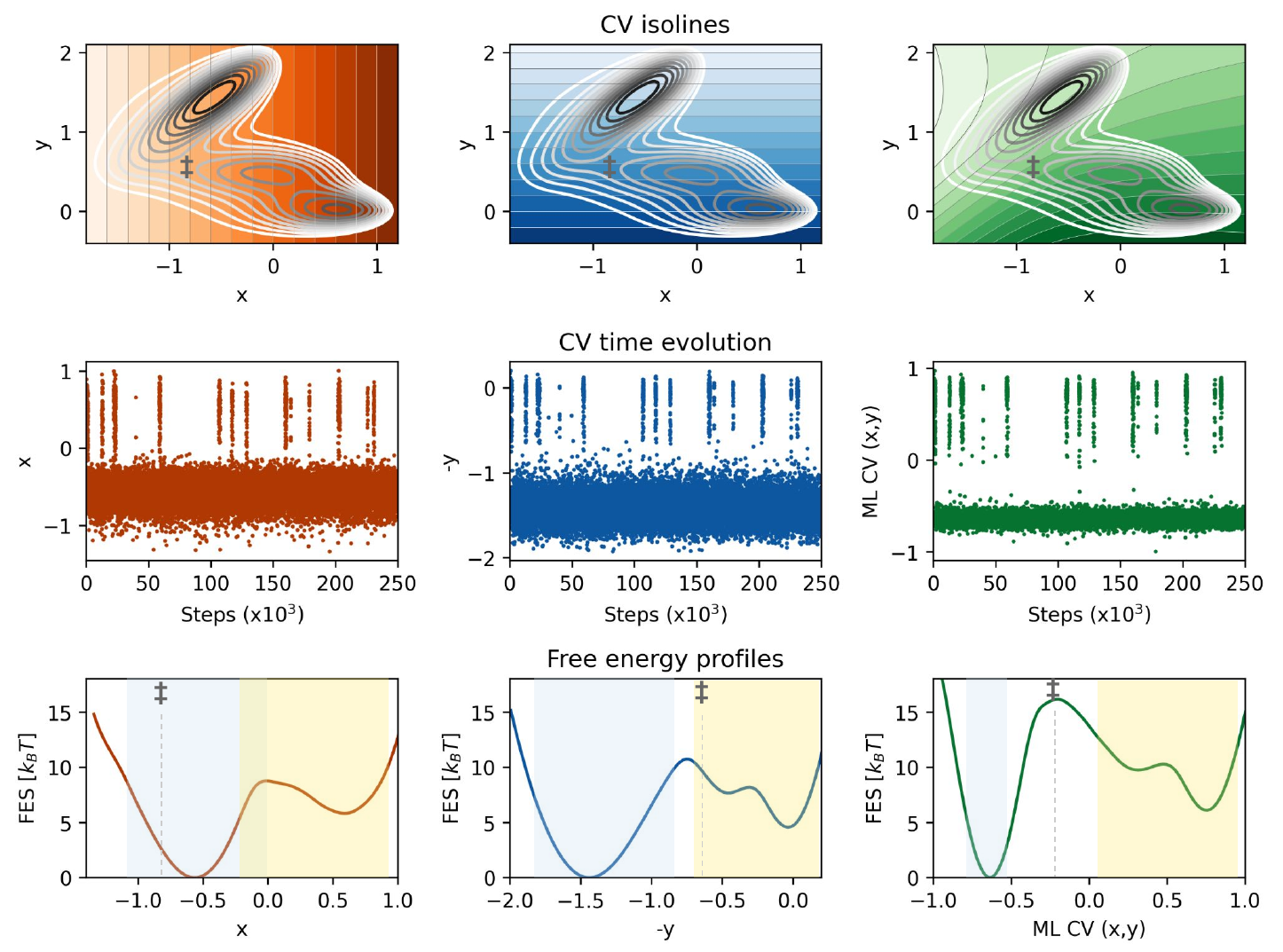}
    \caption{Comparison of different collective variables for a toy model system. 
    Each column represents a different collective variable: $x$, $y$, and a non-linear combination of them  $\text{MLCV}(x,y)$. The first row reports the isolines of each CVs superimposed on the potential energy surfaces of the toy model.
    The second row contains the time evolution of the CVs for a long MD trajectory, and the third row reports the free energy profiles reconstructed along the CVs. 
    The regions corresponding to the metastable states are highlighted by colored shadows and the position of the transition state ($\ddagger$) to allow one to understand whether they are distinct in this projection.}
    \label{fig:cvs}
\end{figure}

In addition, to be effective in the context of enhanced sampling, this function must perform dimensionality reduction. In fact, most biasing schemes work best with few CVs. In addition, CVs should be able to distinguish different metastable states and encode slow modes that hinder sampling. In practice, the latter aspect implies the ability to distinguish transition states from metastable states as well. 
To reflect on the quality of different choices of CVs, we can consider a toy model as shown in Fig.~\ref{fig:cvs}, where we compared three different choices of CVs, respectively the $x$ coordinate (first column), the $y$ coordinate (second column), and finally a nonlinear CV, function of $x$ and $y$. 
The first row shows the isolines of the CVs, which inform us about the direction in which the potential is pushing, which is along the gradient of the CV and thus orthogonal to the isolines.
The second and third rows show the time series of each CV and free energy profiles along them. 
From all these quantities we can observe that the three CVs are progressively better: $x$ is a clearly suboptimal choice, as the projection is not able to completely distinguish even the metastable states.
The variable $y$ is able to do so but does not clearly distinguish the transition state.
Finally, the nonlinear CV is able to distinguish the latter as well.
    
\paragraphtitle{CV construction}. Traditionally, the determination of CVs has mostly been guided by chemical and physical intuition, choosing a small set of descriptors.  
For example, if the process under investigation involves the formation or breaking of chemical bonds, one might identify some relevant distances or functions of them. 
Otherwise, in the case of biological processes, a measure of distance (e.g., RMSD) from a reference (e.g., experimental) configuration has often been employed. 
However, as one can easily imagine, this approach is bound to fail as the complexity of the systems studied increases and as more degrees of freedom are involved in the process. 
It suffices to think of chemical reactions catalyzed by highly dynamic environments, where the catalytic active site undergoes important changes on the same timescales of the reactions. 
To address these challenges, a shift from a physical to a more data-driven approach has started in the last decades. 
These approaches rely on statistical and machine learning techniques to automatically identify CVs from the high-dimensional MD data. 
Data-driven approaches analyze patterns, correlations, and features in the data without prior assumptions about the system. 
Early attempts tried to combine primitive descriptor-based CVs with linear mathematical operations, such as Principal Component Analysis~\cite{joliffe2016pca} (PCA) to find the high variance modes or Time-Lagged Independent Component Analysis~\cite{Perez-Hernandez2013} (TICA) to find the slowly decorrelating ones. 
More recently, there has been a proliferation of methods based on deep learning and neural networks, whose greater expressiveness has made it possible to study increasingly complex systems.  

\subsection{Ingredients of a data-driven approach}
\label{sec:cvs_ingredients}

We now discuss the different ingredients needed to build a machine-learning CV, which are: 
    \begin{enumerate}
    \itemsep0em 
        \item \sansbf{Representation of the system}.
        The input of the model function for the CV should be some featurization of the system. 
        For example, we can start from (a subset of) raw atomic positions or characterize each configuration with a set of physical descriptors. 
        \item \sansbf{Model function}.
        The CV is expressed as a model function $\textbf{s}_\theta(\textbf{x})$ that depends on some learnable parameters $\theta$ that can be optimized to achieve a specific task. 
        Such a function can be, for example, as simple and interpretable as a linear combination or more complex as a feed-forward or a graph neural network.
        \item \sansbf{Data (acquisition)}.
        The data from which to learn and their quality heavily influence the effectiveness of CV research. 
        Typically, they are produced using MD simulations, although they can also be generated artificially. 
        \item \sansbf{Learning objective}.
        To optimize CVs on a given dataset, we need an objective function that mathematically expresses the required task.
    \end{enumerate}
    
As you can already guess from this brief summary, these ingredients cannot be considered independent of each other.
For example, different architectures are designed to work with different representations, and the information that can be extracted from the data naturally depends on their quality. 
In the following, we discuss each of these four ingredients in detail.

\subsubsection{Representation of the system}
\label{sec:cvs_representation}

    The first ingredient of a data-driven approach is the choice of how to represent the system, which is the input to our model function. 
    The choice of representation is guided by multiple requirements: it should be informative enough to describe the physics of interest without leaving out relevant degrees of freedom, but at the same time, it should be computationally efficient and preserve the necessary symmetries. 

    \begin{figure}[b]
        \centering        \includegraphics[width=0.65\linewidth]{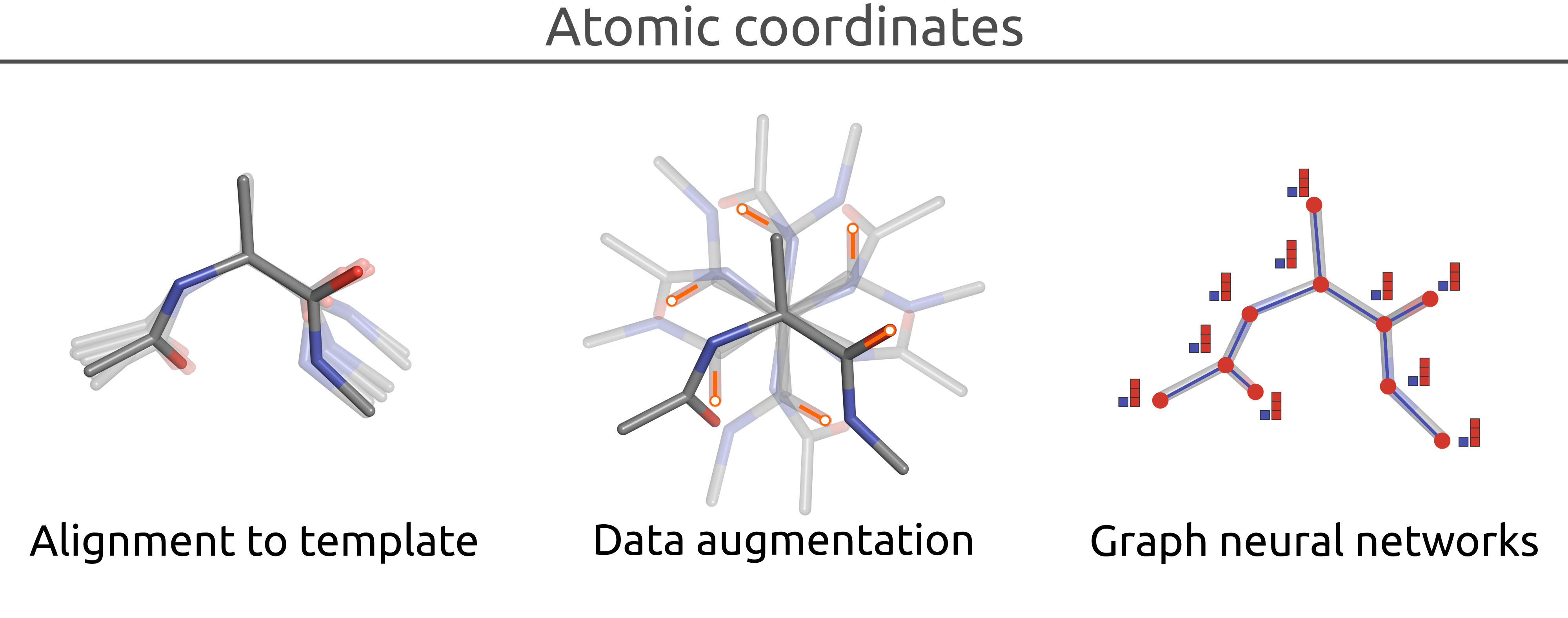}
        \caption{Different ways of using the atomic coordinates as inputs of the ML-based CVs while preserving the necessary symmetries.}
        \label{fig:coordinates}
    \end{figure}
    
\paragraphtitle{Atomic coordinates}.
One possible strategy is to directly use the raw Cartesian coordinates of the atoms as input for our model function.
This is certainly the most natural and direct choice, and in principle, all atoms can be included in the representation. 
Yet, in practice, one typically performs a preliminary pruning of the atoms to reduce the overall computational costs and decrease noise, taking advantage of the fact that some atoms can be harmlessly neglected, e.g., non-reactive hydrogen atoms or carbons in rigid aromatic structures.
One problem with using Cartesian coordinates is that they do not preserve symmetries. 
For this purpose, there are different available strategies (see Fig.~\ref{fig:coordinates}):
    \begin{itemize}
        \itemsep0em
        \item Make an alignment of the system to a reference template. 
        This performs a roto-translation that minimizes the distance to the reference (this can be done directly in PLUMED with \verb|FIT_TO_TEMPLATE| or also in Python). 
        This procedure is feasible where there is a rigid structure and no reactive processes occur that change the topology of the system. 
        \item Learning symmetries implicitly through data-augmentation: by this, we mean the process of artificially generating new data from existing data by applying symmetry operations (e.g., random rotations).
        This can also be applied to ensure permutational invariance.
        \item Recent neural network architectures, such as graph neural networks~\cite{duval2024hitchhikers}, are built to be invariant or equivariant under certain symmetry group transformations, encoding the symmetries directly into the architecture.
        However, this comes at a much higher cost compared with other approaches. 
    \end{itemize}

\begin{figure}[h!]
        \centering        \includegraphics[width=0.65\linewidth]{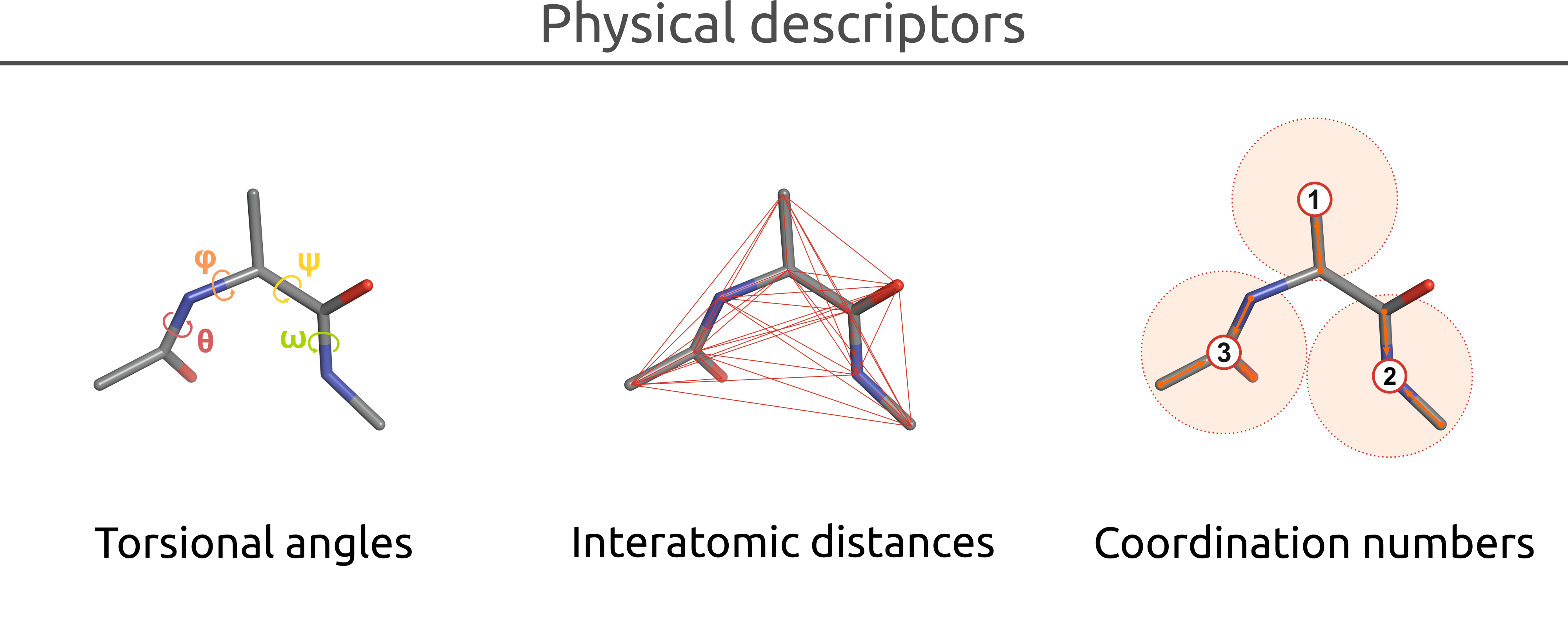}
        \caption{Examples of different physical descriptors.}
        \label{fig:descriptors}
    \end{figure}

\paragraphtitle{Physical descriptors}. 
   As an alternative to using atomic coordinates, a popular choice is to characterize the system through a set of physical descriptors. 
   This provides an initial coarse-grained description of the system, thus allowing the bias to target the degrees of freedom of interest.
   Furthermore, it is a practical way to impose the required symmetries and inject some prior knowledge into the model. 
   For example, interatomic distances are readily invariant under translational and rotational operations, and coordination numbers are also invariant with respect to permutations of identical atoms, a property that is required, for example, when considering solvent interactions or the presence of equivalent reactive atoms in a chemical reaction.
   More complex and abstract functions can be used to characterize the system, like the eigenvalues of adjacency matrices~\cite{pietrucci2011graph, Raucci2022, yang2024sulfur} or structure factor peaks~\cite{Karmakar2021}. 
   Short-range descriptors, such as symmetry functions or smooth overlap of atomic positions~\cite{bartok2013representing} (SOAP), which are widely used for machine learning potentials, can also be employed. 
   However, they are not widely used for biological applications because of their high computational cost. 
   Moreover, a good CV does not necessarily require a perfect and complete representation; simpler and less expensive descriptors can also lead to significant acceleration. 
   Here, we briefly highlight some possible choices (see Fig.~\ref{fig:descriptors}) alongside their definition in PLUMED:

\begin{itemize}
\itemsep0em
    \item \textit{Interatomic distances} (\verb|DISTANCE|) are typically a good representation to describe conformational changes in biomolecules.
    \item \textit{Contacts} (obtained by applying a continuous cutoff-based switching function to distances) may be a better choice to describe a chemical reaction, as they better resemble the on-off nature of chemical bonds.
    \item \textit{Coordination numbers} (\verb|COORDINATION|) provide a simple permutationally invariant description, useful for when equivalent atoms may participate in a reaction or to account for solvent interaction. 
    \item \textit{Dihedral angles} (\verb|TORSION|) can be used to describe the conformational equilibria of proteins, although they should be transformed to sin/cos to obviate their periodicity. 
\end{itemize}

It is important to note that the same information can be provided by different descriptors, although some of them may be easier to interpret. 
In addition to the type, the number of descriptors also requires some observations. 
Although neural networks allow the use of large sets of features, for enhanced sampling applications it is best to filter them to achieve a trade-off between their computational cost and the information they can bring to the CVs.
Generally, a good compromise is in the order of hundreds of descriptors.
Here, we outline some possible strategies for filtering the descriptors:

    \begin{itemize}
        \itemsep0em
        \item A simple way is to look at the distribution of the features in the different states in the system or their changes during the trajectory.
    Descriptors that always have the same value will hardly bring much information to the model. 
    On the other hand, those whose value significantly changes across the different states will likely be more informative.
    \item More complex protocols relying on information theory (like AMINO~\cite{ravindra2020automatic}), as well as genetic algorithms~\cite{hooft021genetic}, can be used to select a subset of descriptors out of a pool of candidates.
    \item Feature selection can also be done iteratively. 
    One can start with a fairly large, agnostic set of descriptors, use it to obtain an initial CV, and then analyze that CV to identify which descriptors are most relevant to the model and exclude those that are redundant and unimportant.
    This can be done using a feature relevance analysis (see Sec.~\ref{sec:sensitivity}) or a sparse approximation of the model using LASSO (Sec.~\ref{sec:sparse_models}).
    \end{itemize}

\subsubsection{Model function}
\label{sec:cvs_model_function}
    In a data-driven approach, the CV is expressed as a model function of the atomic positions $\textbf{s}_\theta(\textbf{x})$ with some learnable parameters $\theta$ that can be tuned to optimize the CV for a specific task. 
    In general, the choice of the functional form is a trade-off between the computational cost, expressiveness, and interpretability of the model.
    The first of these elements is directly related to the time required to perform the underlying mathematical operations to combine the model inputs into the output.
    The other two are more subtle to define. 
    By expressivity, we refer to the ability of our model to extract information from the data, especially when that information is hidden as a complex non-linear function of the descriptors. 
    On the other hand, by interpretability, we refer to the possibility of obtaining information from the optimized model that can help us understand the studied phenomenon. 
    In the following, we discuss different choices for the architectures of the ML-based CVs.
    
    \paragraphtitle{Linear models} lie on one end of this range, being cheap and readily interpretable. 
    The coefficients' values serve as a direct measure of the relevance of each input feature.
    However, this comes at the cost of limited expressivity and requires significant pre-processing in terms of the input features to work properly, i.e., we need to already have a linearly separable space. 
    A few well-known examples of linear techniques are principal component analysis~\cite{joliffe2016pca} (PCA), linear discriminant analysis~\cite{Mendels2018, Piccini2018} (LDA), and time-lagged independent component analysis~\cite{Perez-Hernandez2013} (TICA). 

\begin{figure}[h!]
    \centering
    \includegraphics[width=0.8\linewidth]{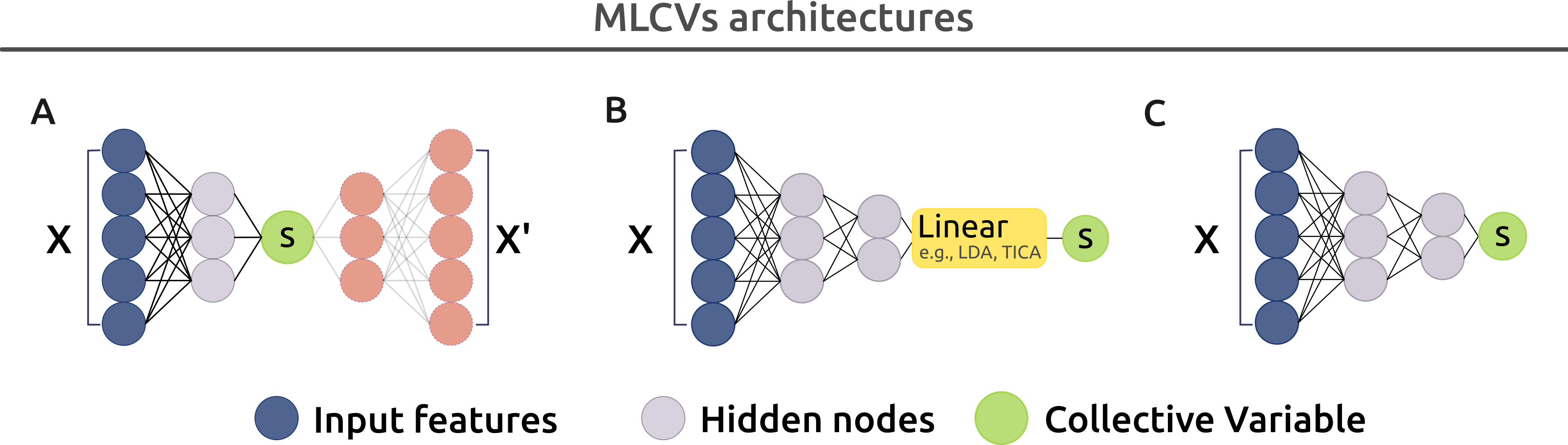}
    \caption{Different architectures for MLCVs based on NNs. \textbf{A}. When using autoencoders, the CV corresponds to the latent space bottleneck of the architecture. \textbf{B}. The NN is used to project the inputs into a lower-dimensional latent space in which a linear statistical method, such as TICA or LDA, is easy to apply. \textbf{C}. The CV space is directly obtained as the output layer of a feed-forward neural network.}
    \label{fig:mlcvs_arch}
\end{figure}

\paragraphtitle{Feed-forward neural networks} (NNs).  
    In this case, the input features are combined via a series of weighted linear combinations and non-linear activation functions through a (deep) layered structure, thus allowing for much greater flexibility and expressivity with respect to the linear counterparts~\cite{bengio2017deep}.  This allows us to combine more complex and numerous descriptors, although at a higher cost and with lower interpretability (which can, however, be recovered by subsequent ad hoc analysis; see Sec.~\ref{sec:interpretability}). In practice, feed-forward NNs have been used in two main ways to obtain CVs.
    \begin{itemize}
        \itemsep0em
        \item NNs can be used as a nonlinear combination of input features, providing a low-dimensional latent space, which is then used as input for a linear technique (e.g., PCA, LDA, TICA...), as schematically depicted in Fig.~\ref{fig:mlcvs_arch} B.  
        The parameters of the NN are optimized based on a loss function that maximizes the score of the underlying linear technique. 
        Examples of this group are DeepLDA~\cite{bonati2020data} (maximizes the separation between the states in the CV space) and DeepTICA~\cite{bonati2021deep} (maximizes the autocorrelation of the CVs). 
        \item  Express to CV space directly as the output of the NN (see Fig.~\ref{fig:mlcvs_arch} C) and encode the optimization in the objective function, as in DeepTDA~\cite{trizio2021enhanced}.
    \end{itemize}
If, in the case of one-dimensional CVs, these two approaches are very similar to each other, they differ more when multiple CVs are to be learned. 
On the one hand, using NN to learn a non-linear featurization allows one to impose important properties on CVs, such as orthogonality between them. 
On the other, this can be a limitation because it constrains the number of CVs, which could be large in some cases. 
Optimizing the CV directly as the output of the network allows for greater flexibility and relaxation of these properties, resulting in lower dimensionality spaces.
    
    \paragraphtitle{Autoencoders} are a particular kind of neural networks, which are composed of two NNs concatenated one after the other, as depicted in Fig.~\ref{fig:mlcvs_arch} A.
    The first one, the \textit{encoder}, maps the high-dimensional input space into a low-dimensional latent space. 
    The second one, the \textit{decoder}, maps the latent space back to the high-dimensional space~\cite{bengio2017deep}.
    Autoencoders can be used in combination with different tasks, such as reconstructing the input or predicting long-time dynamics. 
    In the context of enhanced sampling, the CV is obtained from the latent space bottleneck (i.e., the output of the encoder), while the decoder is used only during the training. 
    AEs have found their way into the Deep-CVs literature by providing the backbone of many methods~\cite{Chen2018,ribeiro2018reweighted,Belkacemi2022,zhang2024combining}. 
    
    \paragraphtitle{Graph neural networks} (GNNs) represent a recent branch of deep learning. 
    In these models, the system is represented by a connected graph in which the atoms correspond to the nodes in the graph, and the edges represent the connections between them~\cite{duval2024hitchhikers}. 
    The properties of the nodes/edges can be either invariant quantities (i.e., they do not change if a symmetry operation such as a rotation is applied) or also equivariant (i.e., they transform with it). 
    Using equivariant features can lead to more expressive and especially data-efficient architectures, although at a much higher computational cost. 
    They also differ from feed-forward NNs in the way they are optimized. 
    A popular option is the message-passing scheme, in which information about a particular node is updated based on the values from neighboring nodes and edges.
    This allows the "messages" to be propagated across the networks. 
    Unlike the other architectures, GNNs have only recently been employed to obtain CVs due to their higher computational cost~\cite{dietrich2024nucleation, zou2024nucleation, vandenhaute2024rare, zhang2024descriptorsfree}.
\vspace{-0.5em}
\subsubsection{Data acquisition}
\vspace{-0.5em}
\label{sec:cvs_data_acquisition}
    As emphasized in previous sections, the foundation of an effective data-driven CV lies in data available for learning. 
    Such data should encompass all relevant metastable states and transition pathways. 
    Any gap in the data can lead to the model extrapolating poorly and producing unreliable results. 
    Furthermore, while the quantity of data plays a role, it is often their quality that determines the model's performance. 
    The type of data collected also dictates the kinds of questions we can explore, such as identifying slow modes (which requires reactive, close-to-equilibrium trajectories) versus high variance ones (also possible with not converged, out-of-equilibrium simulations).
    All these considerations lead to the critical question: how can we obtain such high-quality data?
    
A general way to address this challenge is to resort to an iterative exploration and refinement approach. 
Initially, more aggressive exploration methods can be employed to identify and map out the relevant states.
Afterward, we can use closer-to-equilibrium techniques aimed at enhancing data quality.
This two-step strategy ensures that the collected data are both comprehensive and reliable, forming a robust foundation for the CV model.
    In the following, we briefly discuss some techniques that can be used for a generic collection of data, both to explore and discover states and to harvest reactive pathways.
    
    \paragraphtitle{Exploratory methods to discover states}
    \begin{itemize}
        \itemsep0em 
        \item \textit{Collective variable-free} methods, such as  \verb|OPES_EXPANDED| with multithermal or multithermal-multibaric ensembles (see Sec.~\ref{sec:opes-expanded}).
        The use of these methods is better than running MD simulations at high temperatures, as it also allows the relevant low-temperature configurations to be sampled, using the high-temperature ones to escape the barriers. 
        Of course, this is only applicable if the transition is temperature-driven (e.g., protein folding), while it is not adequate if raising the temperature causes other processes to take place that are not the ones of interest (e.g., in the case of protein-ligand binding, at high temperature, the protein unfolds).
        \item \textit{Generic CVs with exploratory sampling methods}: one can use generic (and, in this sense, suboptimal) CVs to explore the space of configurations.  
        To do this, it is appropriate to use exploratory methods, such as \verb|OPES_METAD_EXPLORE|, or to use them in combination with \verb|OPES_EXPANDED| (e.g., multithermal) to alleviate the suboptimality problem (see Sec.~\ref{sec:opes_explore} and ~\ref{sec:opes-expanded}). 
        These CVs can be physics-suggested quantities, such as the RMSD for protein folding or host-guest distance for ligand binding, but also CVs constructed to blindly explore possible reactions, e.g., based on graph connectivity~\cite{Raucci2022}. 
        Furthermore, also high-variance dimensionality reduction techniques, such as those based on autoencoders, can be used in an iterative scheme to explore the configurational space and find the metastable states (see Sec.~\ref{sec:cvs_autoencoders}).
    \end{itemize}
    
    \paragraphtitle{Methods to harvest transition pathways}
    \begin{itemize}
        \itemsep0em 
        \item \textit{OPES-Flooding} can be used as a tool to collect paths out of a state, with the advantage that it does not require knowing the arrival state. 
        \item \textit{Transition Path Sampling} (TPS) is a well-known technique for sampling reactive paths by performing a Monte Carlo in the space of trajectories. 
        This requires knowing the start and end state, and especially to use other software outside PLUMED.
        \item \textit{Metadynamics of paths}~\cite{mandelli2020metadynamics} exploits concepts from both path sampling and biased simulations to harvest unbiased transition trajectories.
        This technique was used to collect data for the training of CVs using a custom implementation of LAMMPS and PLUMED~\cite{mullender2024effective}.
    \end{itemize}
Finally, we also mention the possibility of artificially generating data to improve CV, either to improve path description (e.g., through geodesic interpolation~\cite{yang2024geodesic}) or to learn a representation in a self-supervised learning framework~\cite{wang2022molecular}.

\subsubsection{Learning objectives}
\label{sec:cvs_learning_objective}
The last ingredient of data-driven CVs is the criteria used to optimize them. 
Various methods have been proposed for the data-driven optimization of CVs, which, from a high-level point of view, can be categorized based on three progressively more specific properties they enforce:
    \begin{enumerate}
    \itemsep0em
        \item \textit{Operate a dimensionality reduction.} 
        As a first basic requirement, a CV should provide a concise representation of the system in a low-dimensional space while preserving as much information as possible, as in Refs.~\citenum{Chen2018,ribeiro2018reweighted,Lemke2019,rydzewski2021multiscale,Belkacemi2022,hradiska2024acceleration}.
        \item \textit{Distinguish the existing metastable states.} 
        Another necessary condition to represent the free energy landscape of the system meaningfully is to map the metastable states to different regions of the CV space without overlaps, as in Refs.~\citenum{Mendels2018,bonati2020data,trizio2021enhanced}
        \item \textit{Reflect the long-term evolution of the system.}
        In other words, this requires the CV to encode the system's slow modes, which are related to the rare transitions between free energy basins, as in Refs.~\citenum{McCarty2017c,Wehmeyer2018b,Chen2019,Hernandez2018,bonati2021deep,wang2021state,shmilovich2023girsanov}.
    \end{enumerate}

Ideally, a good CV should meet all three requirements. 
A representative example is the \textit{committor function}~\cite{weinan2010transition}, which, for the transition between two states, measures the probability of ending up in one state instead of the other. 
This function is one-dimensional and, by construction, can both distinguish states and describe the transition across the free energy barrier. In fact, several methods aim to learn the committor function, either directly or via surrogate objectives~\cite{ma2005automatic,jung2023machine,evans2023computing,kang2024computing,frohlking2024deep,france2024data}.

Of the three properties listed above, the third one (encoding the slow modes) is the most desirable in an enhanced sampling context because it ensures that the procedure accelerates the variables that hinder the sampling, resulting in an effective speedup of our simulations. 
Unfortunately, this is also the most difficult to obtain because, in general, such information can be extracted from data that include reactive trajectories obtained from simulations that are not too far from equilibrium. 
As a result, the data-driven approach to CVs may seem to be a chicken-and-egg problem since to get good CVs, you need good sampling, but to get good sampling, you already need good CVs.
To find a way out of this problem, a successful strategy has been to use surrogate criteria (such as enforcing only one of the first two conditions) to optimize CVs. 
Sometimes, this can yield CVs that are already effective in promoting transitions and converging the simulations. 
Alternatively, if such CVs are not optimal, they may still gather new data that can be used in a second round to apply more refined approaches and obtain better CVs. 

    \paragraphtitle{Learning settings and data}. 
    From the perspective of the learning process, the three objectives listed above can be achieved with different learning schemes, each requiring different \textit{types} of data~\cite{bonati2023mlcolvar}. 
    In general, the goal of simple dimensionality reduction is associated with unsupervised learning, which operates on unlabeled datasets, thus applicable to all data from MD, whether obtained near equilibrium or not.
    On the other hand, if one wants to ensure that the metastable states of the system are clearly distinguishable in CV space, one can use classification-based approaches (supervised learning), which require labeled datasets. 
    For example, this may be a set of configurations of the different basins or even belonging to the transition state and labeled accordingly. 
    Finally, methods for encoding the slow modes of the system typically require a dataset that covers the relevant part of the phase space, including reactive paths. 
    In this case, one can speak of forecasting learning or extraction of variables associated with long autocorrelation times. 
    Finally, there is also the possibility of mixing different approaches, which has been recently proposed to make the best out of different data types we may have at hand without wasting precious and often costly simulation data.
    This can be done by following a multi-task learning approach, in which the CV model is optimized to minimize, at the same time, different objective functions evaluated on different (types of) datasets.

\subsubsection{The computational pipeline with \texttt{PLUMED} and \texttt{mlcolvar}}

\begin{SCfigure}[][h!]
        \centering
        \includegraphics[width=0.7\linewidth]{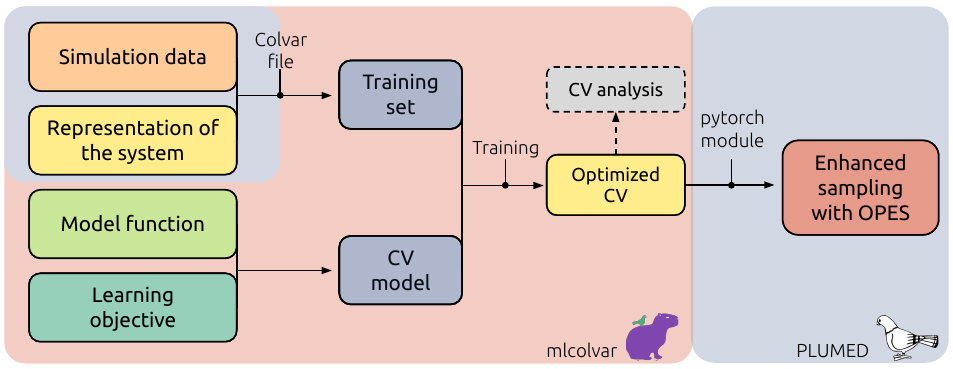}
        \caption{Computational pipeline of machine learning collective variables. 
        This starts with data acquisition (and choice of the representation), which is accomplished via \texttt{PLUMED}. Then, we move to \texttt{mlcolvar} for the training of the MLCVs, where we optimize a given CV model on this training data. Finally, the CV is exported and loaded back into \texttt{PLUMED} to perform enhanced sampling simulations.}
        \label{fig:pipeline}
    \end{SCfigure}

    To conclude this section,  we contextualize the ingredients into the general workflow (schematized in Fig.~\ref{fig:pipeline}) for computing MLCVs with the \verb|mlcolvar| package and their application for enhanced sampling simulations with \verb|PLUMED|.
    
\paragraphtitle{1. Data acquisition - \texttt{[PLUMED]}}:
    The starting point is the collection of the training data and the calculations of physical descriptors, where needed. 
    Although the featurization step can be performed with different libraries, it is appropriate to use PLUMED to ensure that exactly the same setup is used during the training and production phases (i.e., when using CV for enhanced sampling).
    Depending on the type of representation one decides to adopt, we should create a PLUMED input file to compute the input quantities (see Listing~\ref{lst:data_acquisition}), either them being physical descriptors (i.e., \verb|DISTANCE|, \verb|TORSION|, \verb|COORDINATION|\dots) or raw atomic coordinates (i.e., \verb|POSITION|, \verb|CELL|).
    Such quantities are then printed on the PLUMED \verb|COLVAR| files and can be computed either on the fly during an exploratory simulation or using PLUMED as a post-processing tool on an existing trajectory, using the \verb|PLUMED_DRIVER| function.
    
    \begin{figure*}[h!]
    \begin{lstlisting}[language=bash, caption={Minimal example of a PLUMED input file for the calculation of physical descriptors, in this case, the sine and cosine of the $\phi$ torsional angle of Alanine dipeptide.}, label=lst:data_acquisition]
# compute phi torsional angle
phi: TORSION ATOMS=5,7,9,15

# compute sin and cos of phi
sinphi: MATHEVAL ARG=phi FUNC=sin(x) PERIODIC=NO    
cosphi: MATHEVAL ARG=phi FUNC=cos(x) PERIODIC=NO    

# print to colvar file
PRINT ARG=sinphi,cosphi FILE=COLVAR STRIDE=100      
    \end{lstlisting}
    \vspace{-2em}
    \end{figure*}   
    
\paragraphtitle{2. CVs optimization - \texttt{[mlcolvar]}}:
    The next step is the definition of the CV model and its optimization based on the collected data, which are entirely done within the \verb|mlcolvar| framework with a few lines of code, as schematically outlined in Listing~\ref{lst:cv_optimization}.
    The \verb|COLVAR| files can be easily imported in Python with \verb|mlcolvar| through some helper functions available in the I/O module of the library that allow to automatically prepare the data in PyTorch datasets for the training and/or dataframes for analyses. 
    Many CV models can then be easily initialized using the CV classes implemented in the library, and the training is done relying on functions from the high-level \verb|lightning| package.
    After the optimization, the CV is converted to the \verb|TorchScript| language to make it usable in PLUMED.

    \begin{figure*}[h!]
    \begin{lstlisting}[language=python, caption={Minimal code snippet for the training of MLCVs from PLUMED data using the \texttt{mlcolvar} library and their deployment for performing enhanced sampling simulations in PLUMED.}, label=lst:cv_optimization]
import torch,lightning
from mlcolvar.data import DictModule
from mlcolvar.cvs import AutoEncoderCV
from mlcolvar.utils.io import create_dataset_from_files

# 1. Import training data (e.g. PLUMED COLVAR files)
dataset = create_dataset_from_files('./COLVAR')
# 2. Create a Lightning datamodule which splits dataset in train/valid
datamodule = DictModule(dataset, lenghts=[0.8,0.2])
# 3. Choose a model and hyper-parameters
cv_model = AutoEncoderCV(encoder_layers=[45,30,15,2])
# 4. Define a trainer object
trainer = lightning.Trainer(max_epochs=1000)
# 5. Optimize parameters
trainer.fit(cv_model, datamodule)
# 6. Compile the model with TorchScript
cv_model.to_torchscript('model.ptc')
    \end{lstlisting}
    \vspace{-2em}
    \end{figure*}

\paragraphtitle{3. Enhancing sampling - \texttt{[PLUMED]}}.
    Having optimized the CV, the final step is to use it in PLUMED to perform enhanced sampling simulations.
    This can be done using the interface provided by the \verb|PYTORCH_MODEL| function from the optional \verb|pytorch| module of PLUMED, as shown in Listing~\ref{lst:enhanced_sampling}.
    After the initialization with a given label (e.g., \verb|mlcv| in the example), a CV object for each output of the model will be created, characterized by a progressive suffix (e.g., \verb|mlcv.node-0|, \verb|mlcv.node-1|\dots), which can be used to drive enhanced sampling simulations with OPES or other CV-based methods. 
    
    \begin{figure*}[h!]
    \begin{lstlisting}[language=bash, caption={Minimal example of a PLUMED input file for the use of MLCVs to perform enhanced sampling simulations.}, label=lst:enhanced_sampling]
# compute phi torsional angle
phi: TORSION ATOMS=5,7,9,15                                   

# compute sin and cos of phi
sinphi: MATHEVAL ARG=phi FUNC=sin(x) PERIODIC=NO              
cosphi: MATHEVAL ARG=phi FUNC=cos(x) PERIODIC=NO              

# import mlcv model
mlcv: PYTORCH_MODEL FILE=model.ptc ARG=sinphi,cosphi          

# setup OPES calculation
opes: OPES_METAD ARG=mlcv.node-0 PACE=500 BARRIER=30                
# print to colvar file
PRINT ARG=sinphi,cosphi,mlcv.node-0,opes.* FILE=COLVAR STRIDE=500\end{lstlisting}
\vspace{-1em}
    \end{figure*}

\subsection{Representative use cases}

In this section, we discuss some typical learning settings, trying to highlight the heuristics behind them and the data they need. Following the same structure used for the OPES variants, we will also comment on their main parameters along with some suggestions to facilitate their setup and references to relevant examples from the literature.

\subsubsection[Unsupervised learning and exploration with autoencoders]{Unsupervised learning and exploration with autoencoders \quad    \linkbox{MLCOLVAR-\href{https://mlcolvar.readthedocs.io/en/latest/notebooks/tutorials/cvs_Autoencoder.html}{AECV}}}
\label{sec:cvs_autoencoders}

    \mybox{\paragraphtitle{Key point}: We can learn a compressed representation that can retain as much information as possible about the structural changes in the system. 
    This approach is unsupervised and can be used to explore configurational space, as well as being applied to all types of data.    \vspace{0.5em}
    }

\paragraphtitle{Description}. 
Compressed representation can be efficiently implemented using autoencoders (AEs).
As mentioned in Sec.~\ref{sec:cvs_model_function}, this class of neural networks is composed of an encoder $\mathbf{s} = E(\mathbf{x})$, which performs the dimensionality reduction from the input space to the CVs \textit{latent space}, and a decoder $\mathbf{x}' = D(\mathbf{s})$, which is trained to reconstruct the input from the value of the CV.
More specifically, AEs are usually trained to minimize the so-called \textit{reconstruction} loss:
    \begin{equation}\label{eq:loss_ae}
        \mathcal{L}_{\text{AE}} = \sum_{i=1}^n \left| \mathbf{x}_i - D(E(\mathbf{x}_i)) \right|^2
    \end{equation}
As a consequence, the AE is pushed to learn a dimensionality reduction that preserves the information about the modes with the largest variance, which may be associated with large penalties in Eq.~\ref{eq:loss_ae}. 
This can be seen as a non-linear generalization of the popular PCA method~\cite{joliffe2016pca}.
Contrarily to PCA, however, AEs can identify nonlinear modes, but these modes are not guaranteed to be orthogonal, which could complicate their physical interpretation (see Sec.~\ref{sec:interpretability}). Furthermore, it should be noted that the loss function does not act directly on the CV space. For this reason, to enforce structure on the latent space, several modifications have been proposed~\cite{Lemke2019,lelievre2024analyzing}.

\begin{SCfigure}[][h!]
    \centering
    \includegraphics[width=0.65\linewidth]{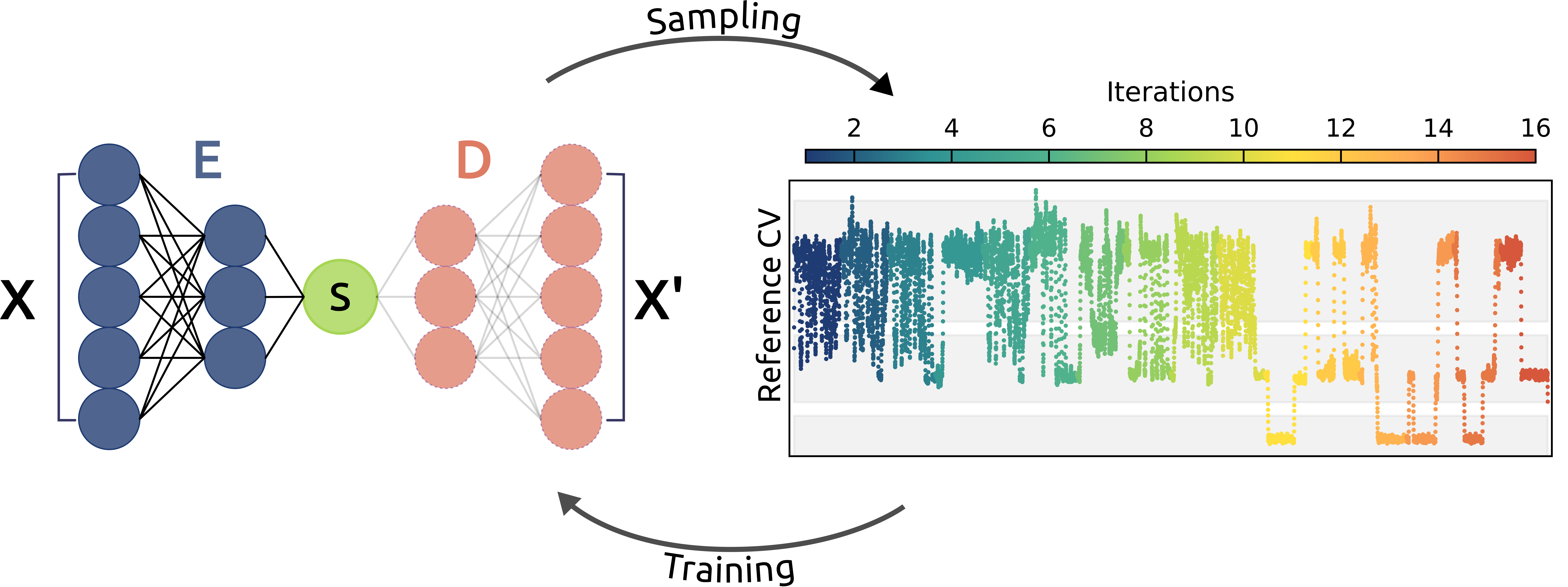}
    \caption{Autoencoders are often employed in iterative procedures alternating cycles of training and sampling.
    This way, for example, one can discover new states through the iterations. The right panel shows the sampling for a three-state toy model from Ref.~\citenum{bonati2023mlcolvar} on a reference CV that can distinguish the three states, highlighted by the gray shaded regions.
    Starting from the topmost basin, two new states are discovered through the iterations, thanks to increasingly better CVs.}
    \label{fig:autoencoders_iterative}
\end{SCfigure}

Due to their unsupervised learning setting, autoencoders have often been used iteratively, alternating cycles of enhanced sampling and CV optimization with the newly obtained data.
In the Molecular Enhanced Sampling with Autoencoders~\cite{Chen2018} (MESA) method, Umbrella Sampling is used to progressively increase the space explored and refine the CV, 
The Free Energy Biasing and Iterative Learning with AutoEncoders~\cite{Belkacemi2022} (FEBILAE) method proposes a similar approach but applies reweighting to biased data during iterations.

\paragraphtitle{Data.} This approach requires, in principle, only a set of configurations for the studied system.
There are, however, a few important caveats.
First, the mode of largest structural change in the provided dataset must correspond to the molecular process you want to enhance, which is the basic assumption of this approach.
Second, the training dataset should ideally include configurations along the full pathway(s) to minimize the risk of the NN interpolating/extrapolating badly in areas unseen during the training, thus hampering the efficiency of the sampling.
Contrarily to the other approaches described below, however, it does not require assigning configurations to states, nor does it require information on the unbiased dynamics of the system.
This is thus a good option for complex processes with many unknown intermediate metastable states that are difficult to identify.

\paragraphtitle{Hyperparameters}:
\begin{itemize}
    \itemsep0em
    \item \textit{NN architecture}: Typically, encoders and decoders are fully connected networks with symmetric architectures, although asymmetric ones have also been proposed (e.g., to improve the interpretability of the CV)
    \item \textit{CV dimension}:  The number of outputs of the encoder (equal to the number of inputs of the decoder) determines the dimensionality of the CV.
    This can be chosen based on the fraction of variance explained ~\cite{Chen2018}.
\end{itemize}

\paragraphtitle{Tips and tricks}:
\begin{itemize}
    \itemsep0em
    \item \textit{Reweighting}: If the training dataset is obtained from biased simulations, the effect of this bias can, in principle, be removed using standard reweighting techniques~\cite{Belkacemi2022}.
    In schemes iterating simulations to enrich the dataset and training, this is technically needed to converge the procedure on a well-defined CV.
    Furthermore, this might help reduce the importance of unphysical configurations that might be sampled due to the presence of the bias.
    Unfortunately, this will also reduce the weight of important high-energy regions, such as transition states, which usually carry information to improve the efficiency of the CV. 
    Whether or how much to weigh the samples is thus a decision that controls a trade-off that must be evaluated case by case.
    \item     \textit{Learning multiple pathways}: it should be noted that an autoencoder with a bottleneck of size one will effectively learn a single direction connecting the data~\cite{bengio2013representation}. If multiple pathways are available, we suggest using a latent space with dimension greater than one or using modifications such as multiple decoders~ \cite{lelievre2024analyzing}.
\end{itemize}

\paragraphtitle{Examples}:
\begin{itemize}
    \itemsep0em
    \item \textit{Trp-cage miniprotein folding} with MESA~\cite{Chen2018} \linkbox{NEST-\href{https://www.plumed-nest.org/eggs/19/065/}{19.065}}: AEs were used to calculate the free energy surface of the (un)folding process of the Trp-cage miniprotein.
    Starting from data obtained from unbiased simulations in its native state, the authors proposed an iterative procedure alternating CV training to umbrella sampling simulations to collect more training data and show that they can reconstruct a FES describing the process in 6 iterations.
    \item \textit{Describing solvent effects} with Permutationally Invariant Networks (PINES) \linkbox{NEST-\href{https://www.plumed-nest.org/eggs/23/024/}{23.024}}: 
    in order to build permutational invariant CVs able to capture solvent effects, the PINES method~\cite{herringer2023permutationally} combines the MESA iterative approach using as input features the permutation invariant vector (PIV)~\cite{herringer2023permutationally}.
    This is used to study the association/dissociation of a NaCl ion pair in water as well as the hydrophobic collapse of a polymer chain.
\end{itemize}

\subsubsection[Distinguishing metastable states with DeepTDA]{Distinguishing metastable states with DeepTDA \quad \linkbox{MLCOLVAR-\href{https://mlcolvar.readthedocs.io/en/latest/notebooks/tutorials/cvs_DeepTDA.html}{DeepTDA}}}
\label{sec:cvs_deepTDA}
    \mybox{\paragraphtitle{Key point}: A good CV should at least distinguish between the metastable states. 
    To enforce such property, we can train CVs as classifiers. 
    This requires a labeled dataset.\vspace{0.5em}}
    
\paragraphtitle{Description}.
    It often happens that one is interested in studying the transition between a set of metastable states that are already known in advance. 
    For example, these states could be the folded and unfolded state of a protein, the reactant and the product state of a chemical reaction, or the bound and unbound state of a molecule to a host system. 
    Sometimes, we have only one state (such as the crystallographic structure of a protein or of a host-guest system), but the other is easily obtained by, for example, raising the temperature. 
    In these cases, one can create a labeled dataset in which each configuration is assigned to a given state. 
    This can be achieved, for instance, by running short, unbiased simulations starting from each state and labeling the configurations accordingly.
    \begin{SCfigure}[][h!]
        \centering
        \includegraphics[width=0.6\linewidth]{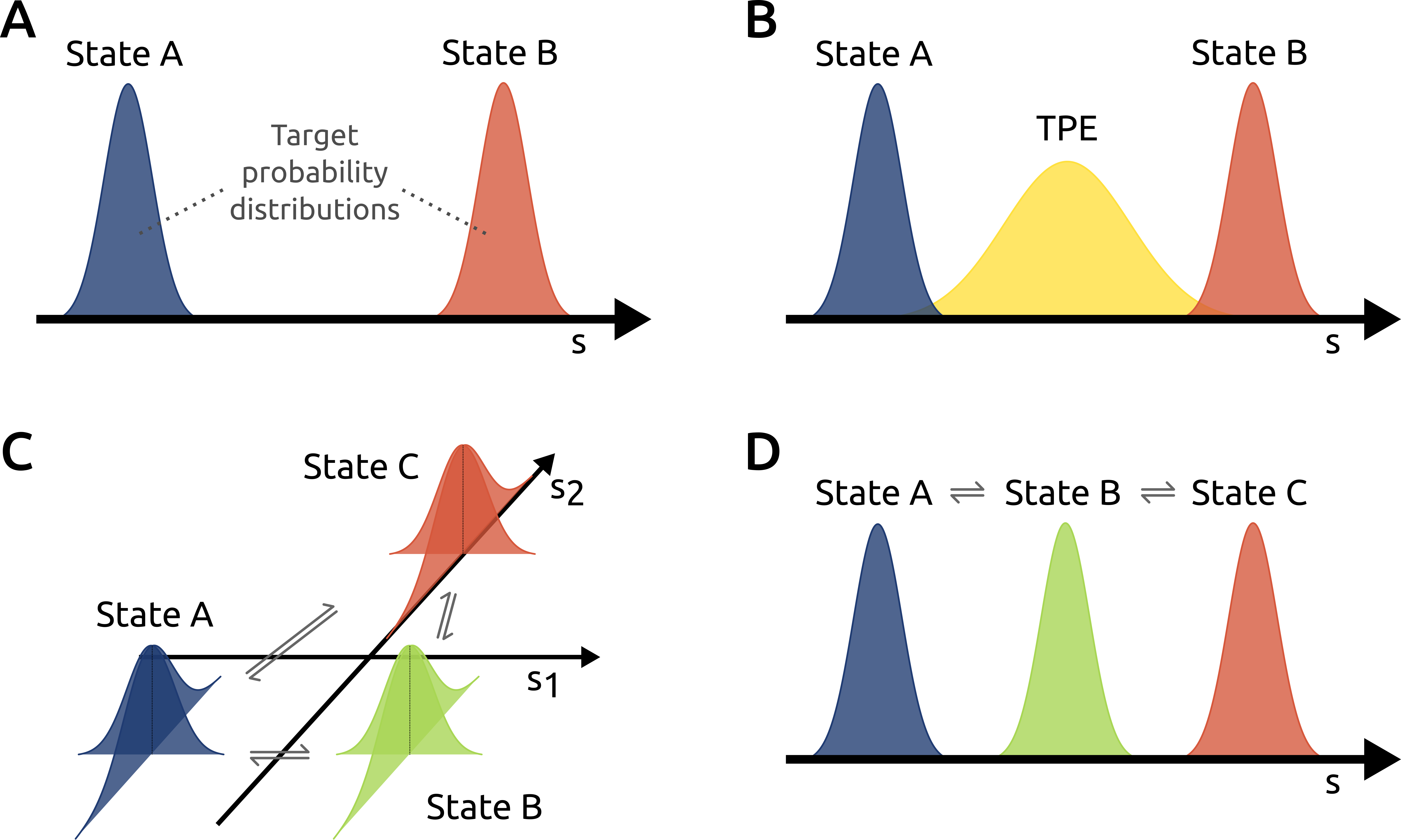}
        \caption{Schematic representation of DeepTDA target distribution composed of series of Gaussians for different scenarios: a two-state system (\textbf{A}), TPI variant to include information on the TPE between two states (\textbf{B)}, general multi-state case with multiple paths, thus requiring multiple CVs (\textbf{C}), multi-state case with linear topology (\textbf{D}), e.g., a chemical reaction with stable intermediates.
        }
        \label{fig:deepTDA}
    \end{SCfigure}
    
    With this kind of data, we can construct a CV with a classification criterion to enforce that the metastable states are well-distinguished when projected in the CV space. 
    Several methods have been proposed to achieve this purpose, and many of them are based on discriminant analysis, such as HLDA~\cite{Mendels2018, Piccini2018} and DeepLDA~\cite{bonati2020data}.
    Here, we present a simpler and more flexible method to achieve this purpose, called deep targeted discriminant analysis~\cite{trizio2021enhanced} (DeepTDA). 
    In DeepTDA \linkbox{NEST-\href{https://www.plumed-nest.org/eggs/21/028/}{21.028}}, the CVs are directly expressed as the output of a neural network (see Fig.~\ref{fig:mlcvs_arch} C), optimized so as to impose that the distribution of the CVs matches a preassigned target distribution in which the states are well distinguished. 
    For instance, the target could be a mixture of Gaussians with diagonal covariances in the CV space and fixed positions and widths. In practice, to achieve this goal, the following loss function is used:
    \begin{equation}\label{eq:deeptda-loss}
        \mathcal{L}_{\text{TDA}} = \sum_k^{N_s} \qty[ \alpha\qty(\boldsymbol{\mu}_k - \boldsymbol{\mu}_k^{tg})^2 + \beta\qty(\boldsymbol{\sigma}_k - \boldsymbol{\sigma}_k^{tg})^2]
    \end{equation}
    For each state $k$ in the summation, the two terms enforces the average $\boldsymbol{\mu}_k$ and standard deviation $\boldsymbol{\sigma}_k$ of the distribution calculated on state $k$ to match the corresponding targets $\boldsymbol{\mu}_k^{tg}$ and $\boldsymbol{\sigma}_k^{tg}$, and the 
    To facilitate the optimization, the two terms are then scaled to have roughly the same magnitude via the hyperparameters $\alpha$ and $\beta$.
    
    Thanks to this simple formulation, DeepTDA can also be easily used to describe multi-state systems.
    In the most general case, given a system with $N_s$ states, one should define a CV space with $N_s - 1$ components in order to account for all the possible transitions between the states.
    However, in many multi-state cases, there's a precise ordering of the states, for example, when a chemical reaction proceeds through a well-defined series of intermediates, e.g., the only possible reaction pathway being A $\Leftrightarrow$ B $\Leftrightarrow$ C. A similar situation is also encountered when the same reactants R can evolve into different products P$_1$ and P$_2$ that cannot interconvert, i.e., P$_1$ $\Leftrightarrow$ R $\Leftrightarrow$ P$_2$.
    In this case, by choosing a one-dimensional target distribution that follows such ordering of the states, DeepTDA can be used to identify a single CV that can describe the whole process, thus improving the computational efficiency and simplifying the interpretability of the results.
    In a similar way, the transition-path-informed variant of the method~\cite{ray2023deep} (TPI-DeepTDA) includes, in an approximated but simple way, information about transition state (TS) configurations by adding a set of configurations representative of the transition path ensemble (TPE) to the training set and mapping them to an additional broader Gaussian bridging the two metastable states.
    
\paragraphtitle{Data.} 
    The method requires a labeled dataset of configurations coming from the different states, which are labeled accordingly. 
    In the case of metastable states, they can be harvested by performing short, unbiased simulations starting from each of the states. 
    Data for the TPI variant can be obtained in different ways, for example, from reactive trajectories from OPES-Flooding~\cite{ray2023deep} or transition path sampling~\cite{zhang2024combining} simulations or even, in the case of proteins, can also be synthetically generated via geodesic interpolation from the initial and final states of the transition~\cite{yang2024geodesic}. 
    
\paragraphtitle{Hyperparameters}:
\begin{itemize}
    \itemsep0em
    \item \textit{Target centers and widths $\{\mu^{tg}, \sigma^{tg}\}$}: the position and standard deviations of the target Gaussians. 
    As a rule of thumb, the target should be chosen to guarantee that different states are neither too close to avoid overlaps between them and to leave enough space in between for the transition states nor too far to not extend too much the region in which the NN is forced to extrapolate.
    Note that what is more important is not the absolute distance between the centers but the relative value compared to the target sigmas.
    For a two-state scenario, we have found the following parameters to work well: $\mu_A^{tg}=-7$ and $\mu_B^{tg}=7$ and $\sigma_A = \sigma_B = 0.2$.
    In the TPI variant, the TPE-Gaussian should be broader and such that it bridges the two states without overlap, e.g., $\sigma_{TPE} = 1$
    \item \textit{$\alpha$ and $\beta$}: Scale the contributions of the center- and sigma-related losses to roughly the same magnitude to have a balanced training. In general, $\beta$ should be larger as the sigma-related loss involves smaller values, e.g., $\alpha = 1$ and $\beta = 100$

\end{itemize}
\paragraphtitle{Tips and tricks}:
\begin{itemize}
    \itemsep0em
    \item \textit{NN outputs}: The only constraint on the architecture is that the number of outputs of the NN determines the number of CVs that can be extracted, and in particular, the number of output nodes corresponds to the number of CVs.
    As a consequence, the targets must also match the size of the CV space.
    \item \textit{TPI variant}: In the TPI variant of the method, it is important to ensure that the TPE dataset only contains reactive configurations and doesn't overlap with the metastable states.
    A good way to check this is to train DeepTDA CV on the metastable states only and use it as a classifier to select the configurations from the TS region.
    
\end{itemize}
\paragraphtitle{Examples}:
\begin{itemize}
    \itemsep0em
    \item \textit{Protein folding} \linkbox{NEST-\href{https://www.plumed-nest.org/eggs/23/009/}{23.009}}: in Refs.~\citenum{ray2023deep, yang2024geodesic}, the DeepTDA method in its transition state aware variant has been applied to the study of the folding of chignolin protein, leveraging information from the folded and unfolded states as well as unfolding trajectories obtained with OPES-Flooding. As input features for the model, the interatomic distances between the $\alpha$-carbons of the protein were used.
    \item \textit{Ligand binding} \linkbox{NEST-\href{https://www.plumed-nest.org/eggs/23/017/}{23.017}}: in Refs.~\citenum{das2023enzyme, das2024correlating}, the DeepTDA method has been used to study the (un)binding process of sugar molecules to the human pancreatic $\alpha$-amylase enzyme in explicit solvent. 
    Two Deep-TDA CVs were trained to describe such a complex system.
    One accounted for the conformational part of the process, using as input features the contacts between relevant atoms in the binding pocket and on the ligand.
    The other described the role of the solvent and was trained using water coordination numbers around the binding site.
    Another more didactical ligand-binding example from the SAMPL5 challenge is presented in Ref.~\citenum{trizio2021enhanced} \linkbox{NEST-\href{https://www.plumed-nest.org/eggs/21/028/}{21.028}}, studying the binding of a small ligand molecule to a calixarene host in an explicit solvent, using the water coordination of a curated set of points as input.

\end{itemize}

\subsubsection[Deep learning the slow modes with DeepTICA]{Deep learning the slow modes with DeepTICA \quad \linkbox{MLCOLVAR-\href{https://mlcolvar.readthedocs.io/en/latest/notebooks/tutorials/cvs_DeepTICA.html}{DeepTICA}}}
\label{sec:cvs_deepTICA}

\noindent\fbox{%
    \parbox{\textwidth}{%
        \paragraphtitle{Key point}: Collective variables should describe the slowest modes in the system. How do you find the slow CVs? By seeking variables characterized by maximum autocorrelation. This requires data of transition between states.\vspace{0.5em}
    }%
} 
\vspace{0.5em}

    \begin{SCfigure}[][h!]
    \centering
    \includegraphics[width=0.5\columnwidth]{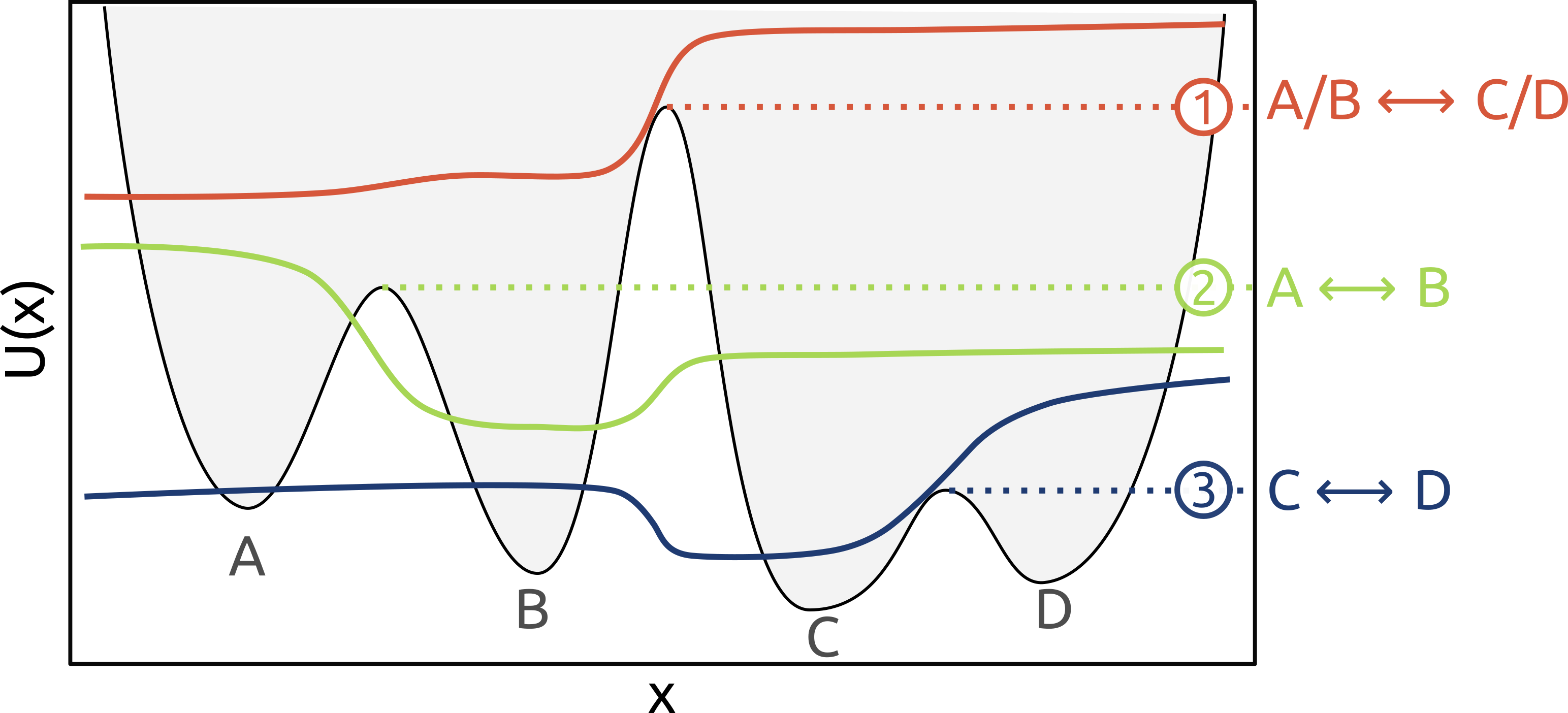}
    \caption{TICA CVs for a toy model system. The black line in the background indicates the potential energy surface of the one-dimensional potential of Ref.~\citenum{Prinz2011}. 
    Above them are the TICA eigenfunctions, which correspond to the maximally autocorrelated modes. These functions have a shape similar to a sigmoid function, which transitions from one value to another near the barrier. Each variable describes progressively faster modes, corresponding to lower barriers.}
    \label{fig:tica-eigenfunctions}
    \end{SCfigure}

\begin{SCfigure}[][t]
    \centering
    \includegraphics[width=0.425\columnwidth]{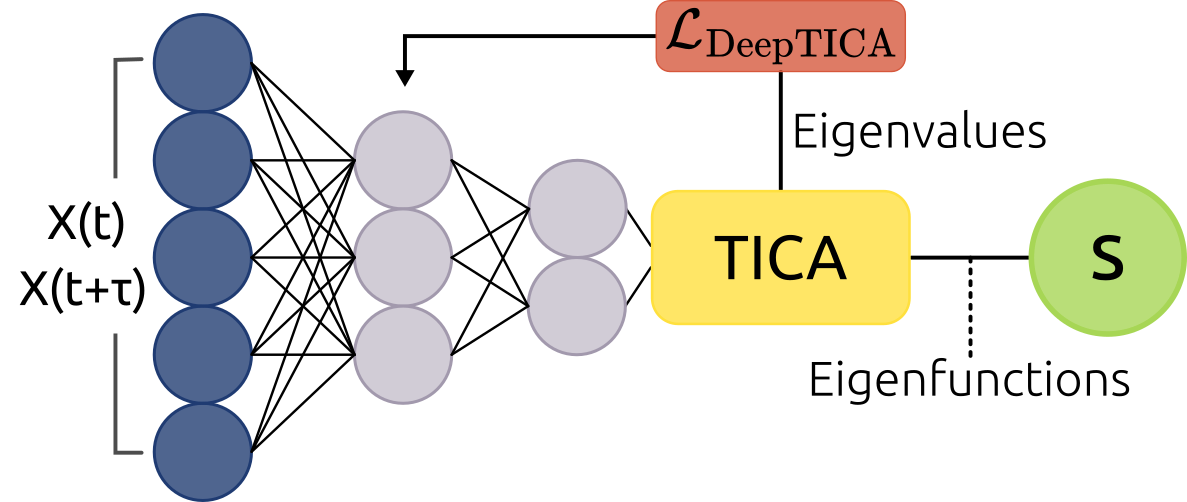}
    \caption{DeepTICA architecture, composed of two parts: 1) a feed-forward NN performing a non-linear featurization of the input features and 2) the TICA transformation, which projects the output of the NN into maximally autocorrelated directions. During training, the parameters of the NN are optimized to maximize the autocorrelation of the final variables by using the sum of squares of the eigenvalues as objective function.}
    \label{fig:deeptica-architecture}
    \end{SCfigure}

\paragraphtitle{Description}. In the framework of the Variational Approach to Conformation Dynamics~\cite{Prinz2011} (VAC), we can relate the eigenfunctions of the transfer operator (which evolves the probability density towards the Boltzmann distribution) to the modes that relax more slowly towards the equilibrium.
In a rare event scenario, these are associated with the rare transitions between metastable states. 
Indeed, these variables have been argued to be natural reaction coordinates~\cite{McGibbon2017}. 
Fig.~\ref{fig:tica-eigenfunctions} shows these functions in the case of a one-dimensional model system, from which we can deduce some interesting features related to their use as CVs.
Each eigenfunction is associated with a different free energy barrier, starting with the highest one, with a functional form that is approximately a sigmoid centered on the barrier.
Consequently, these variables are able not only to classify states but also to accurately describe the transition along free energy barriers.
How can we find such variables for a generic system? We can exploit a variational principle that states that these functions are characterized by maximum autocorrelation. Here in particular we use the DeepTICA method to achieve this goal. Its architecture is composed of two parts (see Fig.~\ref{fig:deeptica-architecture}): 
\begin{enumerate}
\itemsep0em 
    \item a non-linear transformation of the input features via a neural network, followed by
    \item a linear transformation that projects the output of the NN into orthogonal combinations that are maximally autocorrelated. This is accomplished with the TICA (Time-lagged Independent Component Analysis) technique. 
\end{enumerate}
The parameters of the network are optimized so as to maximize the TICA eigenvalues $\lambda_i^{(tica)}$, which are related to the implied timescale $t_i$ measuring the decay time of the i-th eigenfunction: $\lambda_i=e^{-\tau/t_i}$ where $\tau$ is the lag-time used for calculating the autocorrelation. 
This maximization can be accomplished by using a loss function composed of the sum of the first $n$ eigenvalues squared:
\begin{equation}\label{eq:deeptica-loss}
    \mathcal{L}_{\text{DeepTICA}}=- \sum_{i=1} ^{n} \lambda_i^2
\end{equation}

\paragraphtitle{Data.} The data points for the training are pairs of configurations separated by a lag-time $\tau$: $\{\mathbf{x}(t),\mathbf{x}(t+\tau)\}$. 
Unlike autoencoders and classifier-based CV, in order to extract the slow modes it is necessary to have data about transitions between states. 
Ideally, we would like to apply DeepTICA to long, unbiased simulations, but this is rarely feasible. 
Therefore, we can follow a two-step protocol, in which we first run an exploratory advanced sampling simulation (either with generic variables or with generalized ensembles, see Sec.~\ref{sec:opes-expanded}). 
From this simulation we can extract CVs with DeepTICA, which we are going to use as collective variables in a new biased simulation. 
To do this, it is necessary to reweigh the trajectory to account for the presence of the external bias potential.
In DeepTICA, this is done by considering an instantaneous time acceleration due to the bias potential $V$:  $dt'_k = e^{\beta V(\textbf{x}_{t_k})} dt$ (in case of a quasi-static bias like OPES-Metad). 
A discussion on the different reweighting schemes is available in Ref.~\citenum{chen2023chasing}.

\paragraphtitle{Hyperparameters}:
\begin{itemize}
    \itemsep0em
    \item \textit{Time-lag}: An important parameter for TICA is the lag-time $\tau$ between pairs of configurations used to calculate the correlation functions. 
    In the case of unbiased MD simulations, the lag-time has a physical meaning to filter out any process shorter than $\tau$. 
    However, in the biased case where the time is accelerated by the presence of bias potential, this is no longer true.
    We thus suggest choosing the lag-time in a range that meets the following criteria: it must be smaller than the autocorrelation time of all the eigenfunctions that are intended to be recovered, but at the same time, it must be large enough so that the eigenvalues are not degenerate, which makes the NN optimization by gradient descent unstable.  
    This interval can be identified by optimizing the models at different lag-times and plotting the eigenvalues as a function of $\tau$. 
    \item \textit{NN outputs}: The only constraint on the architecture is that the number of outputs of the NN determines the number of CVs that can be extracted, and in particular, it must be greater than or equal to the number of eigenvalues to be optimized. 
    \item \textit{Loss function}: If one wants to optimize many eigenvalues simultaneously, it might be worth trying different ways of combining the eigenvalues in the loss function, such as maximizing only the smallest eigenvalue rather than the sum of them (which favors the largest). 
\end{itemize}

\paragraphtitle{Tips and tricks}:
\begin{itemize}
    \itemsep0em
    \item While the initial biased simulation does not have to be fully converged and this scheme can work with only a few transitions, it is better to use it in combination with a method that converges quickly to a quasi-stationary bias (e.g., OPES-Metad instead of OPES-Explore), as otherwise the signal may be hidden by the bias potential oscillations and the exponential acceleration of times.
    \item Even by reweighting the trajectory, what is learned are the slowly decorrelating modes of the biased dynamics towards the Boltzmann distribution, and not those of the natural system. 
    For example, if a variable has been accelerated by the addition of the bias potential, the signal of that variable in the slow modes may be weaker. 
    Since the main interest here is to enhance the sampling by identifying bottlenecks and accelerating them, this can be easily addressed by biasing the DeepTICA CVs in addition to the previous biasing potential. 
    This can also be achieved by using a static bias obtained at the end of the first simulation without the need to optimize a multidimensional one. 
    \item If your objective is also to recover the unbiased modes, starting from a simulation done with the generalized ensembles (e.g., multicanonical simulation) instead of biasing specific CVs allows one to extract variables that are closer to those of the natural system. 
\end{itemize}

\paragraphtitle{Related methods}.
\textit{Time-lagged AutoEncoders (TAE)}~\cite{Wehmeyer2018b} are autoencoders that construct a compressed representation capable of predicting the $\textbf{x}_{t+\tau}$ configuration from the $\textbf{x}_t$ one instead of merely reconstructing the inputs at a given time. Similarly, the \textit{Variational Dynamics Encoder (VDE)}~\cite{Hernandez2018} method uses a time-lagged Variational AutoEncoder together with an additional term in the loss function that maximizes the autocorrelation of the latent space.
It is worth noting that both TAEs and VDEs tend to learn a mixture of slow and maximum variance modes~\cite{Chen2019capabilities}, unlike TICA-based methods, which learn only the slow decorrelating ones. On the other hand, time-lagged autoencoders do not have the orthogonality constraint on the CVs, so they can compress the latent space even further, which could be useful if there are many slow modes that need to be accelerated. 

\paragraphtitle{Examples}:
\begin{itemize}
    \itemsep0em
    \item \textit{Chignolin folding} \linkbox{NEST-\href{https://www.plumed-nest.org/eggs/21/039/}{21.039}}: in Ref.~\citenum{bonati2021deep} the DeepTICA method has been used to study the folding of the chignolin mini-protein in a blind way. 
    Initial sampling was done with a \verb|OPES_EXPANDED| multicanonical simulation to avoid choosing a CV, which allowed observing a few folding-unfolding events.
    Using these trajectories, a DeepTICA CV was constructed, initially using all the distances between heavy atoms (4278) and then reducing this number to about 200 through a sensitivity analysis procedure (see Sec.~\ref{sec:sensitivity}) to identify in a data-driven way the most relevant input features.
    The DeepTICA CV was then biased in addition to the multicanonical potential, resulting in a 20x improvement in the transition rate and allowing the free energy profiles to be accurately converged over the entire considered temperature range.  
    
    \item \textit{Membrane permeation}: In Ref.~\citenum{muscat2024leveraging}, the DeepTICA approach was used to study membrane binding mechanisms and recover binding affinities of drug-like molecules, validated against experimental data. The initial simulation was an \verb|OPES_EXPANDED| multicanonical simulation, and the system was characterized by a set of (minimum) distances between the molecule and groups of atoms in the membrane. This approach effectively captured the multidimensional nature of membrane insertion, resulting in a CV that could automatically account for molecular orientation, insertion depth, and interactions with different lipid components.

    \item \textit{GPCR ligand binding kinetics}:  a variant of a time-lagged autoencoder based on the RAVE~\cite{ribeiro2018reweighted}/PIB~\cite{Wang2019} was proposed to determine the unbinding kinetics of small molecule drugs~\cite{lamim2020combination}.
    The authors used an initial unbiased simulation to select a maximum of 20 descriptors using AMINO~\cite{ravindra2020automatic} and train a first CV, which was then used in a metadynamics simulation and re-trained on the biased data after reweighting.
    This final CV was used in conjunction with an analytical CV describing the ligand's solvation in an infrequent metadynamics protocol to determine unbinding kinetics.
    
\end{itemize}

\subsubsection[Combining different approaches with Multi-Task CVs]{Combining different approaches with Multi-Task CVs \quad \linkbox{MLCOLVAR-\href{https://mlcolvar.readthedocs.io/en/latest/notebooks/tutorials/adv_multitask.html}{MultiTask}}}

\mybox{\paragraphtitle{Key point}: Enforcing multiple objectives, possibly evaluated on different datasets, allows us to exploit all the available data and obtain more robust CVs. \vspace{0.5em}
}

\paragraphtitle{Description}. In many scenarios, the process of data acquisition through iterations of training and (biased) simulations generates multiple datasets containing different types of information. These could be labeled datasets, which are used to distinguish the states, as well as unlabeled ones coming from biased simulations. How can we combine all of them together? We can use a multi-task approach, in which a single CV is optimized via a loss function defined as the sum of multiple objectives $\mathcal{L}_k$ evaluated on (possibly different) datasets $D_k$.

\begin{SCfigure}[][h!]
    \centering
    \includegraphics[width=0.4\linewidth]{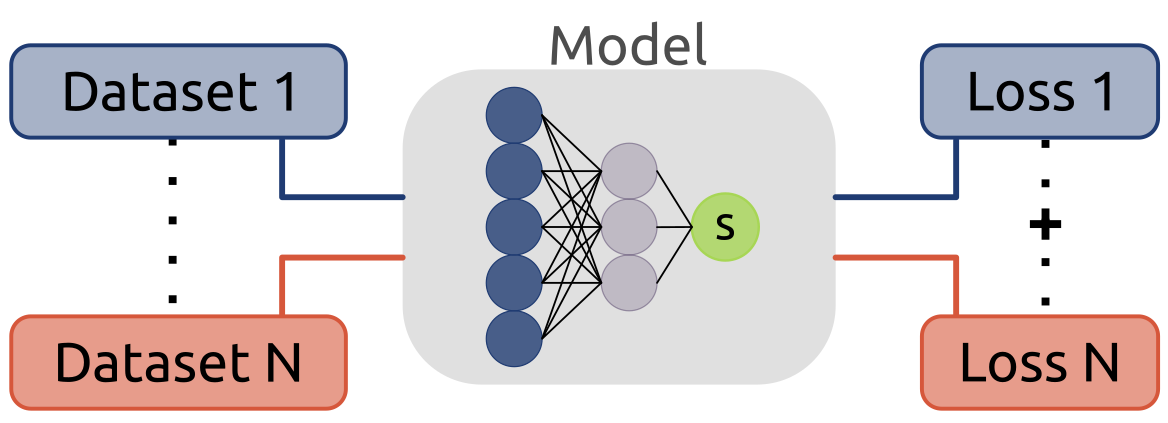}
    \caption{Schematic representation of the working principle of multitask collective variables, in which $N$ dataset, eventually of different type (e.g., labeled and unlabeled), are used to optimize a single model function by minimizing a linear combination of $N$ different loss function.}
    \label{fig:multi_task_cvs}
\end{SCfigure}

In particular, in \texttt{mlcolvar}, it is possible to add to the CV's main loss function $\mathcal{L}_1$ one or more auxiliary loss functions evaluated on separate datasets (Fig.~\ref{fig:multi_task_cvs}). Within this multi-task strategy, the optimized loss is given by
\begin{equation}\label{eq:multitask-loss}
    \mathcal{L}(D_1, \dots, D_K) = \mathcal{L}_1(D_1) + \sum_{k=2}^K \alpha_k \mathcal{L}_k(D_k)
\end{equation}
where $\mathcal{L}_1$ and $D_1$ are the main loss function and dataset, and $\alpha_k$, $\mathcal{L}_k$, and $D_k$ are a weighting coefficient and the loss and dataset of the auxiliary tasks, respectively.

In the example of a semi-supervised approach~\cite{bonati2023mlcolvar,zhang2024combining}, we could set $\mathcal{L}_1$ and $D_1$ to the AutoEncoder loss $\mathcal{L}_{\text{AE}}$ (Eq.~\ref{eq:loss_ae}) and the unlabeled dataset, respectively, while introducing an auxiliary reconstruction loss $\mathcal{L}_2 = \mathcal{L}_{\text{TDA}}$ (Eq.~\ref{eq:deeptda-loss}). This can also be seen as regularizing the latent space learned by the autoencoder leveraging the information on the known metastable states in $D_2$ and that they should be distinguished by the CV. 
Furthermore, in addition to being data-efficient, multi-task strategies have been shown to often result in more accurate and robust NN models in the ML literature.

\paragraphtitle{Data.} With a multi-task strategy, one can mix data of different types to train a single CV. This allows more freedom during the data acquisition process.
Nevertheless, the quality of the data must remain high to obtain good results, and the recommendations given in the sections above remain valid.

\paragraphtitle{Hyperparameters:}
\begin{itemize}
    \itemsep0em
    \item \textit{Main and auxiliary loss functions:} The main hyperparameters is the choice of the loss functions.
    \item \textit{Auxiliary loss coefficients}: The $\alpha_k$ coefficients in Eq.~\ref{eq:multitask-loss} can be used to tune how much weight to give to a particular objective or dataset during training, for instance, to assign higher importance to higher-quality data or to prioritize certain objectives (e.g., force two known metastable states to be distinguished).
    In the absence of any relative importance, coefficients can be tuned to assign equal importance to all loss functions -- often through short trial and error runs -- by selecting values for which the gradients of the loss functions with respect to the NN parameters have equal orders of magnitude.
\end{itemize}

\paragraphtitle{Tips and tricks:}
\begin{itemize}
    \itemsep0em
    \item \textit{Main and auxiliary tasks}: Note that the choice of the main task has meaningful consequences on the properties of the CV model. For example, training a 2-dimensional DeepTICA CV with an auxiliary DeepTDA loss is different than training a 2D DeepTDA CV with an auxiliary loss that maximizes the autocorrelation (Eq.~\ref{eq:deeptica-loss}) because the CV's components will be orthogonal only in the first case. Instead, the auxiliary loss functions could be interpreted as regularization terms for the chosen main CV.
    \item \textit{Single-dataset multi-task strategy}: while the framework in \texttt{mlcolvar}, in its general form, can accommodate multiple datasets of different types, multi-task strategies can also be useful when working with same-type datasets.
    Encodermap~\cite{Lemke2019}, for instance, is a method working on unlabeled datasets that mixes the standard autoencoder reconstruction loss (Eq.~\ref{eq:loss_ae}) with a loss function that encourages the CV to maintain distance relationships between data points.
\end{itemize}

\paragraphtitle{Examples:}

\begin{itemize}
    \itemsep0em
    \item \textit{Combining states labels and potential energies}: An example of single-dataset multi-task strategy was employed in Ref.~\citenum{sun2022multitask}, in which a CV was trained to simultaneously capture information on the assigned metastable state and potential energy of each configuration in order to improve the robustness of the NN model.
    \item \textit{Combining data from MD and TPS simulations}: in Ref.~\citenum{zhang2024combining}, a multi-task framework was used to optimize a CV starting from two different datasets: a set of transitions between two states obtained with TPS and a set of labeled configurations for each of the states. 
    An autoencoder-like CV was optimized with a semi-supervised approach, in which the reconstruction loss was evaluated on the unlabeled TPS data while the TDA loss was optimized on the labeled data. 
    The resulting variable was used to improve the shooting of the TPS simulations and to obtain the free energy through OPES. 
\end{itemize}

\subsection{Interpreting and evaluating ML-based CVs}
\label{sec:interpretability}
One of the most important aspects when it comes to machine learning-based CVs is their evaluation and interpretation.
This is critical not only to be able to gain physical insight from the data but also to understand why the procedure works or not.
This is a very important issue in the use of machine learning methods, and many strategies are available.
What we want to do here is simply to offer some practical advice in the case of machine-learning CVs for enhanced sampling and to suggest some useful tools for this purpose, hoping that it can be a starting point to later use even more advanced tools borrowed from machine learning theory. 

\subsubsection{Correlating the CV with physical descriptors}
\begin{SCfigure}[][h!]\centering
    \includegraphics[width=0.55\columnwidth]{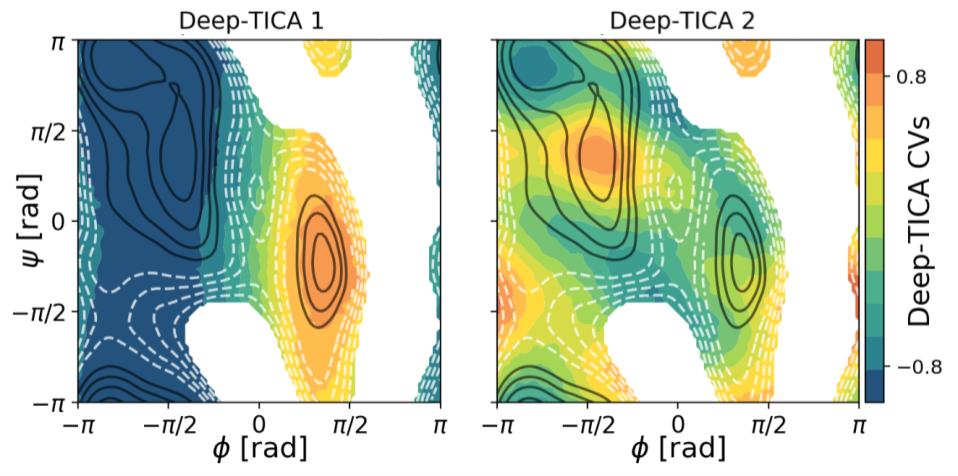}
    \caption{Example of MLCV isolines in a physical space, reproduced from Ref.~\citenum{bonati2021deep}.
    Each plot shows the isolines of the two CVs extracted with DeepTICA for alanine dipeptide in the space of the two torsional angles $(\phi,\psi)$~\cite{bonati2021deep}.
    They are calculated by binning the space and taking for each bin the mean value of the CV for the points within. 
    To facilitate understanding, we superimposed the free energy surface, where metastable states are higlighted with black solid curves.
    From this analysis, we can learn that the DeepTICA 1 CV is associated with the transition between the two free energy basins separated by a barrier around $\phi \sim 0$, while DeepTICA 2 with the faster transition between the two states in the left basin.  }
    \label{fig:ala2-deeptica-isolines}
\end{SCfigure}

Once an ML-based CV is trained, the first question one asks is, what did the CV learn? 
To answer this question, we can try to relate it to some physical quantities which we might know to be relevant to the process under study. 
This can be done in several ways, such as making a scatter plot between the CV and descriptors to find correlated quantities. 
An effective visual procedure is to plot the mean value of the CV $\textbf{s}$ conditioned with respect to physical descriptors $\textbf{d}$. 
This can be done both against a single feature and also in a plane, which allows visualization of CV isolines in a physical space. The latter allows to understand the CVs in terms of the transition between different metastable states, as shown in Fig.~\ref{fig:ala2-deeptica-isolines}).

\subsubsection{Controlling the degree of extrapolation}
\begin{figure}[h!]\centering
    \includegraphics[width=0.85\columnwidth]{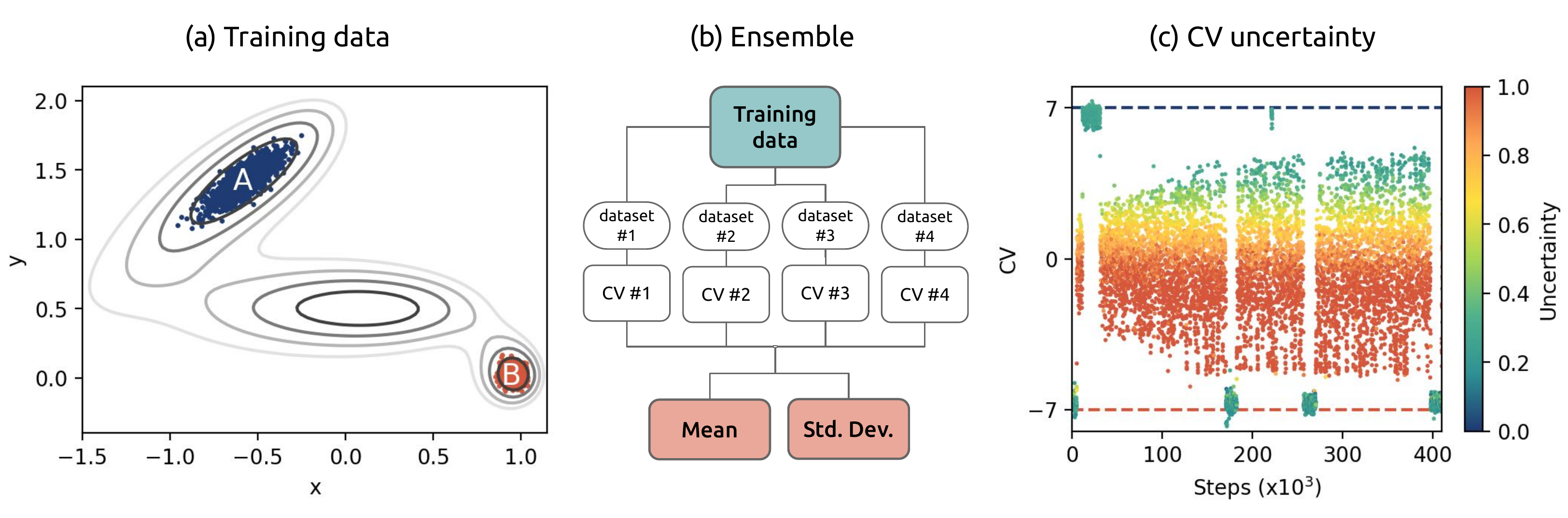}
    \caption{Measuring CVs uncertainty on a toy system (3 state Muller-Brown potential~\cite{bonati2023mlcolvar}). (a) We then train an ensemble of CVs with DeepTDA, using only states A and B (b).
    One of these CVs is then used to enhance sampling, the time trajectory of which is shown in (c), where each point is colored with the uncertainty calculated from the ensemble. 
    We observe that there are three states: at the extremes, we find training states A and B, for which the uncertainty is low, while in the middle, we find a metastable state that is associated with high uncertainty. 
    This is an indicator that we have found a region of space that was not contained in the training set and that we can improve the CV by adding new data from this state. }
    \label{fig:uncertainty}
\end{figure}

It is well known that neural networks are capable of interpolating in the region where data are present in the training set but are not as good at extrapolating into unknown regions. 
That is why it is important to evaluate the degree of extrapolation to know whether one can trust the resulting CV or even to understand whether one is in a region of phase space that was not known.
This could be used, for example, to detect whether a new metastable state was discovered, which might then be included in the CV optimization (see Fig.~\ref{fig:uncertainty}).
One way to assess the degree of extrapolation is to train an \textit{ensemble} of models and use the standard deviation of the predictions as a proxy for uncertainty. 
The rationale is that in the regions similar to the training set, all models will extrapolate similarly, while in far-ranging regions, each might take different directions, a signal that the CV is no longer robust. 
The models should be trained starting from different initializations of the weights and using different splits of the training dataset (e.g., via bootstrap or k-fold splitting).

\subsubsection[Features relevance via sensitivity analysis]{Features relevance via sensitivity analysis  \quad \linkbox{MLCOLVAR-\href{https://mlcolvar.readthedocs.io/en/latest/notebooks/tutorials/expl_features_relevances.html}{Sensitivity}}}
\label{sec:sensitivity}
Being able to identify which features contribute the most to the CV is crucial in interpreting it.
One way to compute feature importance is to perform a sensitivity analysis using the partial derivatives method. 
In this method, we start by computing the gradients of the CV with respect to the input features $x_i$ over a dataset 
$\{\mathbf{x}^{(j)}\}_{j=1} ^N$, that is $\sigma_i\frac{\partial s}{\partial x_i}\Bigr|_{\mathbf{x}^{(j)}}$
where $\sigma_i$ is the standard deviation of the i-th feature, which is needed if the dataset is not standardized.
Once we have the sensitivity per sample, we can calculate the mean (absolute) relevance $r_i$ as 
\begin{equation}
    r_i = \frac{1}{N} \sum_j \left|{\frac{\partial s}{\partial x_i}\Bigr|_{\mathbf{x}^{(j)}}}\right| \sigma_i
\end{equation}

\begin{itemize}
\itemsep0em
    \item This provides information about an average sensitivity. 
    Nonetheless, the CV could give importance to different features in different regions of the configurational space (e.g. in different basins or in the transition region). 
    To assess this, one could restrict the mean to a subset of (labeled) samples.
    \item Once we have identified the features with the highest sensitivity, we can visualize their distribution for the training dataset to see if they are indeed relevant to the process of interest (e.g., whether they distinguish metastable states).
    \item This technique can also be used to reduce the number of inputs by first training a CV with the complete set of features and then training a new one with only a subset of the most relevant ones.
\end{itemize}

\begin{SCfigure}[][h!]\centering
    \includegraphics[width=0.4\columnwidth]{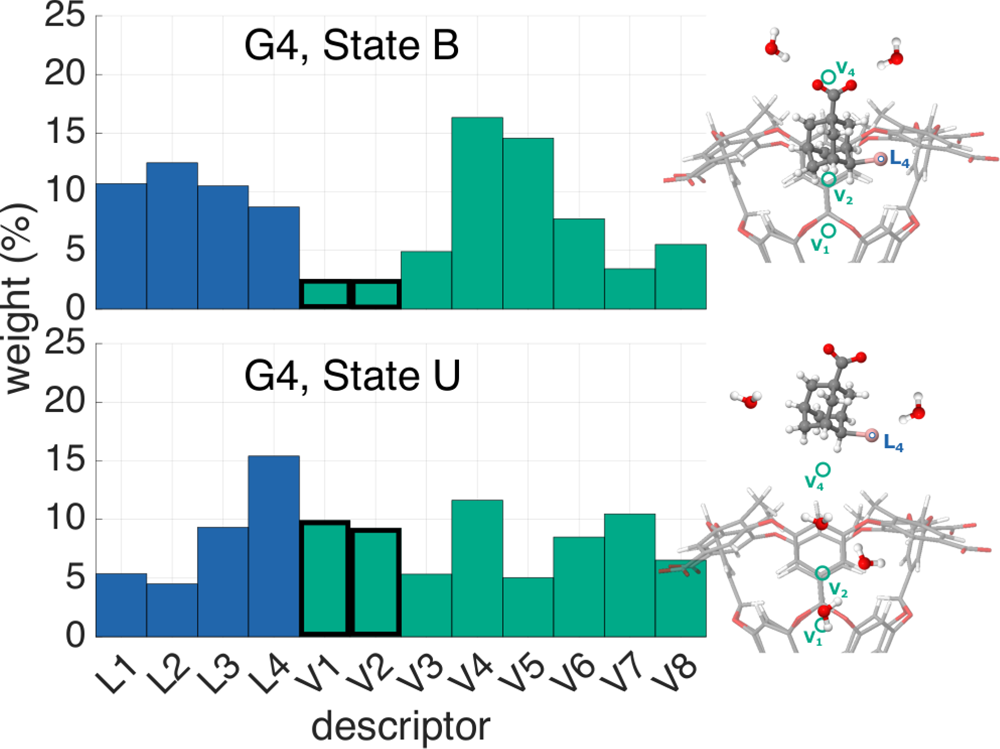}
    \caption{Example of feature importances for a DeepLDA CV applied to a binding process between a calixarene host and a small molecule. 
    The importances are calculated separately for the bound state (B) and the unbound state (U), restricting the set of samples on which the average is performed.
    In this case, the sensitivity of the descriptors is indeed different in the two states, which may provide additional insights into the process. 
    Specifically, the $V_1$ and $V_2$ descriptors describing the presence of water molecules within the host pocket were high in the unbound case and low in the bound case, suggesting a different role in the two states. 
    Image reproduced from Ref.~\citenum{Rizzi2021}.}
    \label{fig:calixarene-ranking}
\end{SCfigure}

\subsubsection[Interpretability via sparse linear models]{Interpretability via sparse linear models \quad \linkbox{MLCOLVAR-\href{https://mlcolvar.readthedocs.io/en/latest/notebooks/tutorials/expl_lasso.html}{Sparse}}}
\label{sec:sparse_models}
One option for interpreting complex, nonlinear models is to approximate them with a linear model: $f_w(\textbf{x}) = \sum_{i=1} w_i x_i $, 
where the coefficients are an immediate measure of the features relevance.
However, to obtain a transparent result, it is necessary that the model is also sparse, i.e., only a few $w_i$ should be different from zero.  
This property can be easily enforced by adding regularizations to the loss function, as in LASSO~\cite{tibshirani1996regression} and Elastic Net~\cite{zou2005regularization}. 
In the following, we briefly describe two scenarios in which sparse linear models (based on LASSO) can be used to characterize the metastable states found by MLCVs and approximate the function represented by a non-linear model with a simpler expression. 

\paragraphtitle{Characterizing metastable states using a sparse classifier}.
In the case of unsupervised methods, whether looking for structure-preserving modes such as autoencoders or slow ones such as in DeepTICA, one often obtains CVs describing transitions between different states. 
The presence of metastable states can be checked by calculating the free energy profile along CVs to find candidates and secondly by running short, unbiased simulations as a verification. 
However, it is often not straightforward to figure out which states are involved. 
For example, in the case of proteins, it is easy to recognize macroscopic changes such as folding, which can also be easily related to known quantities, while it is more complicated to recognize more localized changes, such as those due to interactions between side chains.
In Ref.~\citenum{novelli2022characterizing}, to answer this question and characterize metastable states in a physically transparent way, we used a sparse classifier to identify a minimal set of features that can distinguish them. 
This can be done in a space of one or more CVs, using free energy basins to label configurations. 
Once the labels are assigned, the linear model is optimized with the logistic loss, which ensures the model can distinguish the states and an L1 regularization term on the weights that enforces the sparsity of the model.
Considering for simplicity the two-state scenario ($\xi  = \pm1$ in the following), this amounts to optimizing the model with the following loss function: 
\begin{equation}
    \mathcal{L} = \frac{1}{N}\sum_{\xi  = \pm1}\sum_{j\in \xi} \log \left[e^{-\xi f_{\mathbf{w}}(\mathbf{x}_{j})} + 1\right] + \alpha\left \Vert \mathbf{w}\right \Vert_{1}
\end{equation}
where the first term represents the standard logistic loss and the second is the L1 regularization.
As an example of the workings, Fig.~\ref{fig:bpti-lasso} shows the classification of the states associated with the first three DeepTICA CVs for the BPTI protein and the selected descriptors. In the following, we provide some notes and suggestions:

\begin{SCfigure}[][b]\centering
    \includegraphics[width=0.7\columnwidth]{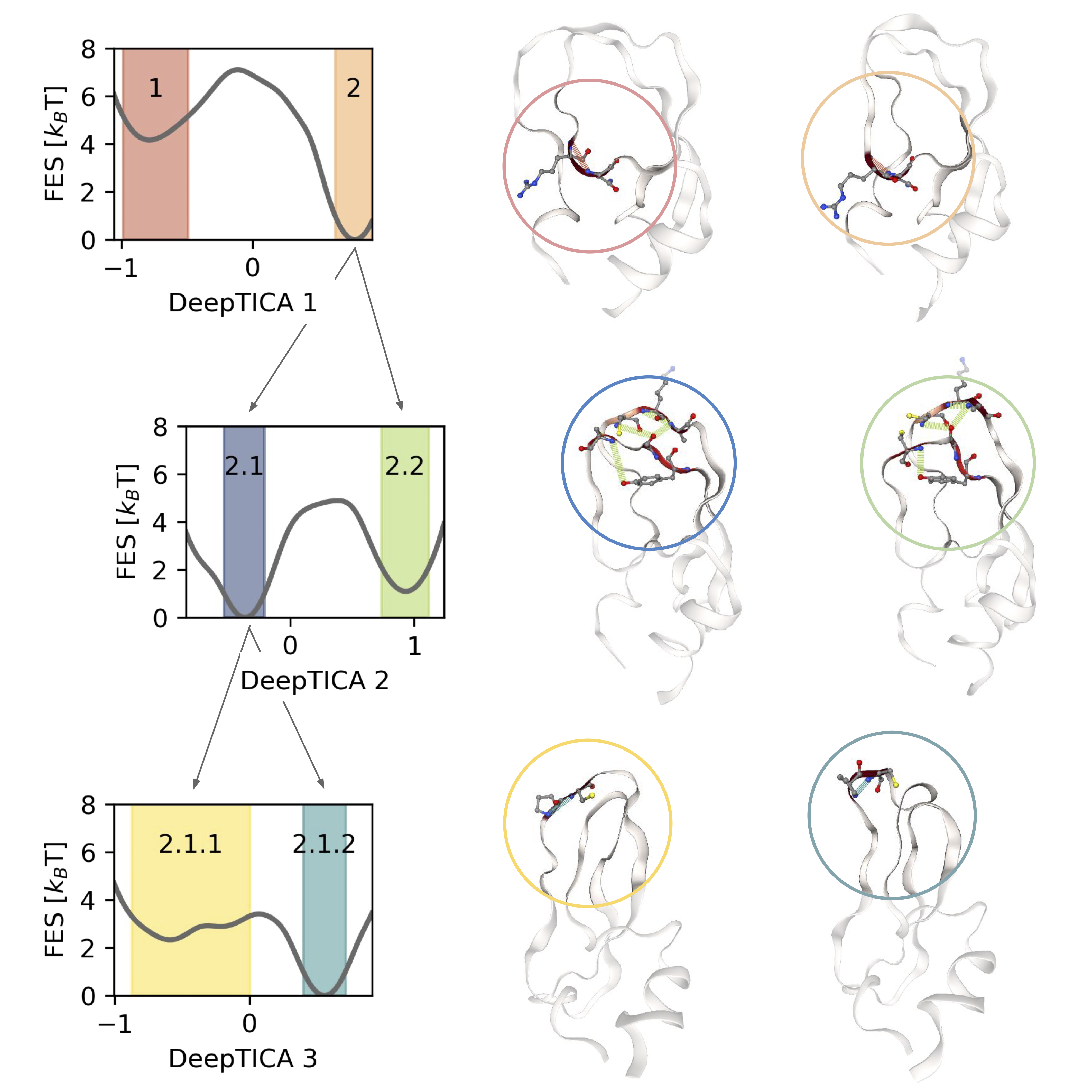}
    \caption{Example of LASSO classification applied to the bovine pancreatic trypsin inhibitor (BPTI) protein. 
    Three CVs were extracted with DeepTICA and used to label states according to free energy (left images). 
    Here, we exploited the hierarchical property of TICAs (see Fig.~\ref{fig:tica-eigenfunctions}), which allows us to divide processes according to their time scales. 
    Highlighted on the right are the descriptors selected by the sparse classifier as important for distinguishing states, and the regions containing them (circles). Reproduced from.~\cite{novelli2022characterizing}. }
    \label{fig:bpti-lasso}
\end{SCfigure}
\begin{itemize}
    \itemsep0em
    \item The value of $\alpha$ should be chosen as high as possible (means a higher sparsity) while maintaining a high classification accuracy. Note that sometimes, the inverse of the regularization strength $C=1/\alpha$ is used.
    This can be made automatic, for instance, by keeping a variable only if it increases the accuracy by at least 1\%. 
    \item In the case of DeepTICA variables, one can also take advantage of the fact that, where there is a clear time scale separation, the CVs take the form of a sigmoid describing the transition from one state to another (see Sec.~\ref{sec:cvs_deepTICA}). 
    This makes it easy to label configurations belonging to metastable states and also to proceed hierarchically, considering first the slowest variable, then the behaviour of the second within each basin identified by the former, and so on (see Fig.~\ref{fig:tica-eigenfunctions}).
    \item The advantage of using sparse linear methods is that we can seamlessly use thousands or more features, automatically selecting the few needed to distinguish the states. 
    However, it should be noted that if there are correlated features, only one of them will be selected, perhaps not the clearest from a physical point of view. 
    Prefiltering the descriptors might thus be useful.
    \item By repeating the analysis using different sets of descriptors, such as distances, dihedral angles, or even more elaborate quantities, one can get information about the states from multiple points of view.  
    \item In the case of $N$ states, instead of using a single classifier, we found it more interpretable to construct $N$ classifiers, each identifying the features that distinguish that state from all others. 
    Note that to improve understanding, it is always possible to limit the analysis between pairs of states.
\end{itemize}

\paragraphtitle{Explaining CVs with LASSO regression}.
Another way to interpret the ML-based CVs is to train a sparse linear model $f_w(\textbf{x})$ that can approximate the more expressive but also cryptic CV optimized with neural networks $f_{NN}(\textbf{x})$~\citenum{zhang2024descriptorsfree}.
The spirit is similar to that of symbolic regression~\cite{jung2023machine}, although easier to optimize because the functional form considered is limited to a linear combination. 
Being a regression problem, this can be achieved with the following loss function:
\begin{equation}
    \mathcal{L} = \frac{1}{N} \sum_j ^N \left \Vert f_{w}(\mathbf{x}_j)-f_{NN}(\mathbf{x}_j)\right \Vert^2+ \alpha\left \Vert \mathbf{w}\right \Vert_{1}
\end{equation}
where the first term leads the regression and the second one enforces again the sparsity of the model.
\begin{itemize}
    \item As before, the value of the regularization strength $\alpha$ must be selected by comparing the accuracy of the regression and the number of selected features. 
    The quality of the approximation can be easily visualized by plotting the output of the two models (i.e., $f_{w}$ and $f_{NN}$) and choosing the first value of $\alpha$ that is able to give a qualitatively correct description of the CV. 
    \item Correlations between variables can be introduced into this scheme by adding to the list of descriptors also the products between descriptors: $\{x_1^2, x_2^2, .., x_1x_2, x_1x_3,.. \}$. 
\end{itemize}

\section{Concluding remarks}
In this chapter, we presented some recent advances in the field of CV-based enhanced sampling. In particular, we discussed the OPES framework for the application of the bias potential and the use of machine learning techniques in CV design. The continuing evolution of these methods, boosted by the new frontiers opened by machine learning, is improving our ability to simulate and analyze complex molecular systems and to understand the fundamental processes that govern them. 
These developments have made enhanced sampling protocols more systematic, expanding their efficiency and scope, and have already been successfully applied to many relevant processes. Nevertheless, physical intuition and system knowledge are still important for their success (e.g., in choosing how to represent the system), and a certain sensitivity is also needed to choose the right tools and combine them. For this reason, we have highlighted the key principles of the methods and provided practical suggestions to guide the reader in designing a successful enhanced sampling study, including those tips, tricks, and details that are often hard to find in papers but are crucial for a successful application. 

\section*{Acknowledgments}
The authors are grateful to Umberto Raucci, Francesco Mambretti, Davide Mandelli, Dhiman Ray, Sudip Das, Peilin Kang, Timothée Devergne, Jintu Zhang, and Simone Perego for helpful discussions and for providing feedback on the manuscript, and to Prof. Michele Parrinello for his guidance in many of the works reported in this chapter.

\clearpage
\section*{References}
\markboth{REFERENCES}{REFERENCES}
\bibliography{main}

\end{document}